\RequirePackage{fix-cm}
\documentclass[a4, twocolumn, epjc3, comma, sort&compress, natbib]{svjour3}
\bibpunct{[}{]}{,}{n}{}{,} % to get "[numbered]" references from natbib

\journalname{Eur. Phys. J. C}

\usepackage{latexsym}
\usepackage{amsmath}
\usepackage{amssymb}
\usepackage{amsfonts}
\usepackage{dsfont}
\usepackage{enumitem}
\usepackage{slashed}
\usepackage{bm}

\usepackage[mathscr,scaled=1.15]{urwchancal}
\DeclareFontFamily{OT1}{pzc}{}
\DeclareFontShape{OT1}{pzc}{m}{it}%
{<-> s * [1.15] pzcmi7t}{}
\DeclareMathAlphabet{\mathpzc}{OT1}{pzc}{m}{it}

\usepackage{color}
\definecolor{purple}{rgb}{0.5,0,0.5}
\definecolor{blue}{rgb}{0.0,0,0.9}
\definecolor{prdblue}{rgb}{0.133,0.118,0.498}
\usepackage[colorlinks=true, pdfstartview=FitV, linkcolor=prdblue, citecolor= prdblue, urlcolor=prdblue]{hyperref}

\usepackage{enumitem}
\usepackage{multirow}%for tabular
\usepackage{subfigure}
\usepackage{textcomp}%for +-

\usepackage{supertabular}
\usepackage{placeins}
\usepackage{epsfig}
\usepackage{graphicx}

\usepackage{dcolumn,booktabs}
\newcolumntype{d}[1]{D{.}{.}{#1}}

\begin{document}

\title{$\,$\\[-6ex]\hspace*{\fill}{\normalsize{\sf\emph{Preprint no}.\ NJU-INP 040/21}}\\[1ex]
Heavy+light pseudoscalar meson semileptonic transitions}
%Contact interaction study of Heavy+light pseudoscalar mesons and their semileptonic decays}

%%\author{Zhen-Ni Xu\thanksref{eZNX,NJU,INP}
%%        \and
%%       Zhu-Fang Cui\thanksref{eZFC,NJU,INP} %etc.
%%       \and
%%       Craig D.~Roberts\thanksref{eCDR,NJU,INP}
%%       \and
%%       Chang Xu\thanksref{eCX,NJU,INP}
%%}

\author{Zhen-Ni Xu\thanksref{eZNX}
        \and
       Zhu-Fang Cui\thanksref{eZFC} %etc.
       \and
       Craig D.~Roberts\thanksref{eCDR}
       \and
       Chang Xu\thanksref{eCX}
}

\thankstext{eZNX}{znxu@smail.nju.edu.cn}
\thankstext{eZFC}{phycui@nju.edu.cn}
\thankstext{eCDR}{cdroberts@nju.edu.cn}
\thankstext{eCX}{cxu@nju.edu.cn}

\authorrunning{Zhen-Ni Xu \emph{et al}.} % if too long for running head

\institute{
School of Physics, Nanjing University, Nanjing, Jiangsu 210093, China %\label{NJU}
           \and
\mbox{$\;$}Institute for Nonperturbative Physics, Nanjing University, Nanjing, Jiangsu 210093, China %\label{INP}
            }

\date{2021 March 28}
%\date{2021 March 02}

\maketitle

\begin{abstract}
A symmetry-preserving regularisation of a vector$\times$vector contact interaction (SCI) is used to deliver a unified treatment of semileptonic transitions involving $\pi$, $K$, $D_{(s)}$, $B_{(s,c)}$ initial states.  The framework is characterised by algebraic simplicity, few parameters, and the ability to simultaneously treat systems from Nambu-Goldstone modes to heavy+heavy mesons.  Although the SCI form factors are typically somewhat stiff, the results are comparable with experiment and rigorous theory results.  Hence, predictions for the five unmeasured $B_{s,c}$ branching fractions should be a reasonable guide.  The analysis provides insights into the effects of Higgs boson couplings via current-quark masses on the transition form factors; and results on $B_{(s)}\to D_{(s)}$ transitions yield a prediction for the Isgur-Wise function in fair agreement with contemporary data.
\end{abstract}

% insert suggested PACS numbers in braces on next line \phi
%\pacs{}

%\keywords{}

\maketitle

%%%%%%%%%%%%%%%%%%%%%%%%%%%%%%%%%%%%%%%%%%%%%%%%%%%%%%%%%%%%%%%%%%%%%%%%%%%%%%%%
%%%%%%%%%%%%%%%%%%%%%%%%%%%%%%%%%%%%%%%%%%%%%%%%%%%%%%%%%%%%%%%%%%%%%%%%%%%%%%%%

\section{Introduction}
\label{Sec:Introduction}
%
%% matrix elements: |Vcd| = 0.218(4), |Vcs| = 0.997(17), |Vub| = 0.00394(36), |Vcb| = 0.0422(8).
%
Pseudoscalar mesons are fascinating for many reasons.  In the light-quark sector, the pion is remarkably light owing to dynamical chiral symmetry breaking \cite{Horn:2016rip}; itself, a corollary of emergent hadron mass (EHM) \cite{Roberts:2020hiw, Roberts:2021xnz}.  Compared with the pion's light $u$, $d$ quarks, the $s$ quark in the $K$ couples more strongly to the Higgs boson.  Thus, the kaon is heavier.  However, if these Higgs couplings are switched off, then the $\pi$ and $K$ are practically identical: they are Nature's most fundamental Nambu-Goldstone bosons.  Consequently, comparisons between pion and kaon properties provide information on empirical expressions of EHM and insights into the constructive interference between Nature's two most basic mass generating mechanisms \cite{Aguilar:2019teb, Chen:2020ijn, Zhang:2021mtn, Roberts:2021nhw, Anderle:2021wcy, Arrington:2021biu}.

These observations are highlighted by studies of the leptonic and semileptonic decays of the $\pi$ and $K$, which measure the scale of EHM, the strength of Higgs boson couplings into quantum chromodynamics (QCD), and the size of the entries in the Cabbibo-Kobayashi-Maskawa (CKM) matrix that generate the associated flavour-changing weak interactions.  In these cases, the CKM matrix elements are $|V_{ud}|$ and $|V_{us}|$.  The former has been measured via the $\pi^+ \to \pi^0 e^+ \nu_e$ transition \cite{Pocanic:2003pf}, leading to  \cite[Sec.\,12]{Zyla:2020zbs}:
\begin{equation}
|V_{ud}| = 0.9739(29)\,.
\end{equation}
The merits of this process lie in the facts that such semileptonic transitions proceed solely through the vector component of the weak interaction and there are no in-medium effects to consider, as there would be with measurements of nuclear transitions.

Semileptonic $K$ decays are similarly used to extract $|V_{us}|$.  In this case, the zero-recoil value of the dominant transition form factor is not unity, so the measurements constrain a product \cite[Sec.\,12]{Zyla:2020zbs}:
\begin{equation}
\label{Vusfp0}
|V_{us}| f_+^{K\pi}(0) = 0.2165(4)\,.
%% = 0.2165 ± 0.0004
\end{equation}
$f_+(0)$ can be calculated using continuum and lattice methods; and an average of modern lattice results yields \cite{Aoki:2019cca}: $f_+^{K\pi}(0) = 0.9706(27)$; hence, $|V_{us}| =0.2231(7)$.

In connection with Eq.\,\eqref{Vusfp0}, an average of continuum predictions \cite{Ji:2001pj, Chen:2012txa, Yao:2020vef} yields
\begin{equation}
f_+^{K\pi}(0) = 0.972(8)\,.
\end{equation}
These Dyson-Schwinger equation (DSE) studies \cite{Eichmann:2016yit, Burkert:2017djo, Fischer:2018sdj, Qin:2020rad} employed the symmetry-preserving leading-order (rain\-bow-ladder, RL) truncation of the continuum bound-state problem \cite{Munczek:1994zz, Bender:1996bb}; and no tuning effort was made.  In combining them here, we exploited the fact that infrared observables are largely insensitive to the momentum dependence of the bound-state kernels owing to the emergence of a gluon mass-scale $m_0 =0.43(1)\,$GeV \cite{Boucaud:2011ug, Aguilar:2015bud, Huber:2018ned, Cui:2019dwv, Roberts:2021nhw}.
%% fp0 = {0.964, 0.98, 0.972}

In the tower of flavoured pseudoscalar mesons, the next states are the $D$ and $D_s$.  The interplays between EHM and Higgs-boson (HB) mass generating effects in these systems are quite different from those in the lighter pseudoscalar mesons.  Naturally, $D$ and $D_s$ would be indistinguishable from the $\pi$ and $K$ if all Higgs couplings were removed; but with the empirical values, the Higgs mechanism alone produces approximately 70\% of their masses and constructive EHM+HB interference generates the remaining 30\%, \emph{i.e}.\ a 70:30 balance.  The analogous results for the $K$ and $\pi$ are 20:80 and 5:95.
Plainly, EHM is subdominant but still significant in $D$ and $D_s$ mesons.

These mass budgets were computed using information from Ref.\,\cite[Figs.\,1.1, 2.5]{Roberts:2021nhw}, including predictions for the active masses of the $c$ and $b$ quarks at $\zeta_2=2\,$GeV, \emph{viz}.\, 1.27\,GeV and $4.18\,$GeV, respectively.  Averages of lattice results produce the following values for these quantities \cite{Aoki:2019cca}: $1.280(13)$, $4.198(12)$.

Precise data are available for $D$ and $D_s$ semileptonic decays \cite{Besson:2009uv, Lees:2014ihu, Ablikim:2015ixa, Ablikim:2018upe, Kou:2018nap, Yuan:2019zfo}.  They provide access to $|V_{cd}|$ and $|V_{cs}|$; albeit, once again, multiplied by soft form factors: \linebreak $f_+^{D_s K}(0)$, $f_+^{D\pi}(0)$ and $f_+^{DK}(0)$.  Lattice-QCD (lQCD) results are available for the last two \cite{Lubicz:2017syv} and continuum results for all three \cite{Yao:2020vef}:
\begin{equation}
\begin{array}{l|lll}
& \ f_+^{D_s K}(0) & \ f_+^{D\pi}(0) & \ f_+^{DK}(0) \\ \hline
\mathrm{continuum} \; \mbox{\cite{Yao:2020vef}} & \ 0.673(40) & \ 0.618(31) & \ 0.756(36) \\
\mathrm{lattice} \; \mbox{\cite{Lubicz:2017syv}} &  & \ 0.612(35) & \ 0.765(31) \\
\end{array}\,.
\end{equation}
A fourth form factor is not typically considered because the accuracy of isospin symmetry means $D^+\to \bar K^0$ and $D^0 \to K^-$ are practically equivalent.

Including the $b$ quark, one has three more pseudoscalar systems: $B_c$, $B_s$, $B$.  In these cases, the HB \emph{vs}.\ EHM+HB mass budgets are 87:13 for the $B_c$ and 80:20 for $B_s$, $B$.  Plainly, the Higgs mechanism is dominant, but the role of EHM is persistently non-negligible.  There are also many more possible semileptonic transitions:
$B_c \to B$ links with $|V_{cd}|$;
$B_c \to B_s$ with $|V_{cs}|$;
$B\to\pi$ and $B_s \to K$ with $|V_{ub}|$;
and
$B\to D$, $B_s\to D_s$, $B_c \to \eta_c$ with $|V_{cb}|$.
The last two open windows onto the third column of the CKM matrix.  However, robust theoretical predictions for all the associated semileptonic transition form factors are difficult to obtain \cite{Gambino:2020jvv}, owing chiefly to the vast separation between the masses of the participating valence-quarks and \mbox{-antiquarks}, and the diversity of competing scales in the processes.

The approach employed in Ref.\,\cite{Yao:2020vef} can be adapted to predicting the leptonic and semileptonic decays of mesons containing a $b$ quark.  In preparing for such a unifying study using realistic quark+antiquark interactions, it is helpful to have a benchmark.  One is provided by calculations made with a symmetry-preserving formulation of a vector$\times$vector contact interaction (SCI) \cite{Roberts:2010rn, Roberts:2011wy}.  Such an approach preserves the character of more sophisticated treatments of the continuum bound-state problem whilst nevertheless enabling an algebraic simplicity; and widespread use has shown that, when interpreted carefully, SCI predictions provide a meaningful quantitative guide, see \emph{e.g}.\ Refs.\,\cite{Wang:2013wk, Segovia:2014aza, Xu:2015kta, Bedolla:2015mpa, Bedolla:2016yxq, Serna:2017nlr, Algarin:2017wmp, Raya:2017ggu, Yin:2019bxe, Gutierrez-Guerrero:2019uwa, Yin:2021uom, Zhang:2020ecj, Lu:2021sgg}.  Thus, with SCI results in hand, one has the means to check the validity of algorithms employed in studies that rely (heavily) upon high performance computing.

This discussion is arranged as follows.
Section~\ref{Sec:SCI} explains our implementation of the SCI, including constraint of the ultraviolet cutoff and determination of values for the interaction-dependent current-quark masses.
Section~\ref{SecSemiLep} details the SCI treatment of semileptonic transitions that proceed via the weak vector vertex, using $D \to \pi^-$ as the exemplar.
Section~\ref{SecSemiLepResults} discusses SCI results for the three $D_{(s)}\to \pi, K$ transitions that are distinguishable in the isospin-symmetry limit, including comparisons with other studies and experiment, where available.
Section~\ref{BWSLT} proceeds with a similar discussion of the seven independent $B_{(c,s)}$ semi\-leptonic transitions, which present many challenges to experiment and theory \cite{Gambino:2020jvv}.
Section~\ref{ESensitivity} considers the question of environmental sensitivity, \emph{i.e}.\ the evolution of semileptonic transition form factors as the Higgs mechanism of current-quark mass generation becomes a more important component of the final-state meson's mass.
Section~\ref{Sec:IWF} reports SCI predictions for the Isgur-Wise function \cite{Isgur:1989ed}, as inferred from the $B_{(s)}\to D_{(s)}$ transitions.
Section~\ref{Epilogue} presents a summary and perspective.

%increasing mass of one quark makes treatments difficult ... even bc systems have very large disparity and bu is terrible
%treatments of light ps mesons and semilep decays available
%few internally consistent + unified treatments across all mass-scales
%which are measured?

\section{Contact Interaction}
\label{Sec:SCI}
When studying the continuum meson bound-state problem, the primary element is the quark+antiquark scattering kernel.  In RL truncation, it can be written ($k = p_1-p_1^\prime = p_2^\prime -p_2$, $k^2 T_{\mu\nu}(k) = k^2 \delta_{\mu\nu} - k_\mu k_\nu$):
\begin{subequations}
\label{KDinteraction}
\begin{align}
\mathscr{K}_{\alpha_1\alpha_1',\alpha_2\alpha_2'}  & = {\mathpzc G}_{\mu\nu}(k) [i\gamma_\mu]_{\alpha_1\alpha_1'} [i\gamma_\nu]_{\alpha_2\alpha_2'}\,,\\
 {\mathpzc G}_{\mu\nu}(k)  & = \tilde{\mathpzc G}(k^2) T_{\mu\nu}(k)\,.
\end{align}
\end{subequations}

The defining quantity is $\tilde{\mathpzc G}$.  Twenty years of study have revealed much about its pointwise behaviour.  A key finding is that, owing to the emergence of a gluon mass-scale in QCD \cite{Boucaud:2011ug, Aguilar:2015bud, Huber:2018ned, Cui:2019dwv, Roberts:2021nhw}, $\tilde{\mathpzc G}$ is nonzero and finite at infrared momenta; so, one may write
\begin{align}
\tilde{\mathpzc G}(k^2) & \stackrel{k^2 \simeq 0}{=} \frac{4\pi \alpha_{\rm IR}}{m_G^2}\,.
\end{align}
QCD has \cite{Cui:2019dwv}: $m_G \approx 0.5\,$GeV, $\alpha_{\rm IR} \approx \pi$.
We follow Ref.\,\cite{Yin:2021uom}, retaining this value of $m_G$, and since the SCI cannot support relative momentum between bound-state constituents, simplifying the tensor structure in Eqs.\,\eqref{KDinteraction} such that, in operation:
\begin{align}
\label{KCI}
\mathscr{K}_{\alpha_1\alpha_1',\alpha_2\alpha_2'}^{\rm CI}  & = \frac{4\pi \alpha_{\rm IR}}{m_G^2}
 [i\gamma_\mu]_{\alpha_1\alpha_1'} [i\gamma_\mu]_{\alpha_2\alpha_2'}\,.
 \end{align}

We implement a rudimentary form of confinement in the SCI by introducing an infrared regularisation scale, $\Lambda_{\rm ir}$, when defining bound-state equations \cite{Ebert:1996vx}.  This device excises momenta less than $\Lambda_{\rm ir}$, so eliminating quark+antiquark production thresholds \cite{Krein:1990sf}.  The usual choice is $\Lambda_{\rm ir} = 0.24\,$GeV \cite{Roberts:2011wy}.

At the other extreme, the integrals appearing in SCI bound-state equations require ultraviolet regularisation.  This breaks the link between infrared and ultraviolet scales that is a feature of QCD.  The ultraviolet mass-scales, $\Lambda_{\rm uv}$, become physical parameters that may be interpreted as upper bounds on the momentum domains whereupon distributions within the associated systems are effectively momentum-independent.  For instance, the $\pi$-meson is larger in size than the $B$-meson; hence, one should expect $1/\Lambda_{\rm uv}^\pi > 1/\Lambda_{\rm uv}^{B}$.  As subsequently explained, this observation leads us to a completion of the SCI through introduction of a scale-dependent coupling \cite{Raya:2017ggu, Yin:2019bxe, Gutierrez-Guerrero:2019uwa, Yin:2021uom}.

{\allowdisplaybreaks
The SCI gap equation for a quark of flavour $f$ is
\begin{align}
\label{GapEqn}
S_f^{-1}(p)  & = i\gamma\cdot p +m_f \nonumber \\
& \quad + \frac{16 \pi}{3} \frac{\alpha_{\rm IR}}{m_G^2}
\int \frac{d^4q}{(2\pi)^4} \gamma_\mu S_f(q) \gamma_\mu\,,
\end{align}
where $m_f$ is the quark's current-mass.  Using a Poincar\'e-invariant regularisation, the solution is
\begin{equation}
\label{genS}
S_f(p)^{-1} = i \gamma\cdot p + M_f\,,
\end{equation}
with $M_f$ obtained by solving
\begin{equation}
%M = m +  \frac{M}{3\pi^2 m_G^2} \,{\cal C}(M^2;\tau_{\rm ir},\tau_{\rm uv})\,,
%M = m +  \frac{M}{3\pi^2 m_G^2} \,{\cal C}^{\rm iu}(M^2)\,,
M_f = m_f + M_f\frac{4\alpha_{\rm IR}}{3\pi m_G^2}\,\,{\cal C}_0^{\rm iu}(M_f^2)\,,
\label{gapactual}
\end{equation}
where
\begin{align}
\nonumber
{\cal C}_0^{\rm iu}(\sigma) &=
\int_0^\infty\! ds \, s \int_{\tau_{\rm uv}^2}^{\tau_{\rm ir}^2} d\tau\,{\rm e}^{-\tau (s+\sigma)}\\
& =
%\overline{\cal C}^{\rm iu}(\sigma) =
\sigma \big[\Gamma(-1,\sigma \tau_{\rm uv}^2) - \Gamma(-1,\sigma \tau_{\rm ir}^2)\big].
\label{eq:C0}
\end{align}
$\Gamma(\alpha,y)$ is the incomplete gamma-function.
In general, functions of the following form arise in solving SCI bound-state equations ($\tau_{\rm uv}^2=1/\Lambda_{\textrm{uv}}^{2}$, $\tau_{\rm ir}^2=1/\Lambda_{\textrm{ir}}^{2}$):
%\begin{subequations}
\begin{align}
%\overline{\cal C}^{\rm iu}_0(\sigma) & =\Gamma(-1,\sigma \tau_{\textrm{uv}}^{2}) - \Gamma(-1,\sigma \tau_{\textrm{ir}}^{2}), \\
%
%\overline{\cal C}^{\rm iu}_1(\sigma)& = \Gamma(0,\sigma \tau_{\textrm{uv}}^{2}) - \Gamma(0,\sigma \tau_{\textrm{ir}}^{2}), \\
%
%\overline{\cal C}^{\rm iu}_2(\sigma) & = \Gamma(1,\sigma \tau_{\textrm{uv}}^{2}) - \Gamma(1,\sigma \tau_{\textrm{ir}}^{2})\,,\\
%
\overline{\cal C}^{\rm iu}_n(\sigma) & = \Gamma(n-1,\sigma \tau_{\textrm{uv}}^{2}) - \Gamma(n-1,\sigma \tau_{\textrm{ir}}^{2})\,,
\label{eq:Cn}
\end{align}
%\end{subequations}
${\cal C}^{\rm iu}_n(\sigma)=\sigma \overline{\cal C}^{\rm iu}_n(\sigma)$, $n\in {\mathbb Z}^\geq$.}

Pseudoscalar ($J^P=0^-$) mesons emerge as quark +anti\-quark bound-states.  They are described by a Bethe-Salpeter amplitude, whose SCI form is \cite{Chen:2012txa}:
\begin{align}
\Gamma_{0^-}(Q) = \gamma_5 \left[ i E_{0^-}(Q) + \frac{1}{2 M_{f g}}\gamma\cdot Q F_{0^-}(Q)\right]\,,
%\Gamma_{0^-}(Q) = \gamma_5 \left[ i E_{0^-} + \frac{1}{M}\gamma\cdot P F_{0^-}\right]\,.
\label{PSBSA}
\end{align}
$M_{f g}= M_f M_{g}/[M_f + M_{g}]$, $Q$ is the bound-state's total momentum, $Q^2 = -m_{0^-}^2$, $m_{0^-}$ is the meson's mass.

The amplitude and $m_{0^-}^2$ are obtained by solving the following Bethe-Salpeter equation: $(t_+ = t+Q)$:
\begin{align}
\Gamma_{0^-}(Q)  & =  - \frac{16 \pi}{3} \frac{\alpha_{\rm IR}}{m_G^2} \nonumber \\
&\times
\int \! \frac{d^4t}{(2\pi)^4} \gamma_\mu S_f(t_+) \Gamma_{0^-}(Q)S_g(t) \gamma_\mu \,.
\label{LBSEI}
\end{align}
Our symmetry-preserving approach implements a di\-men\-sional-regularisation-like identity \cite{Chen:2012txa}:
\begin{equation}
0 = \int_0^1d\alpha \,
%\big[ {\cal C}^{\rm iu}(\omega(M_f^2,M_{\bar g}^2,\alpha,Q^2))\\
\big[ {\cal C}_0^{\rm iu}(\omega_{fg}(\alpha,Q^2))
%
%&& \quad + \, {\cal C}^{\rm iu}_1(\omega(M_f^2,M_{\bar g}^2,\alpha,Q^2))\big],
+ \, {\cal C}^{\rm iu}_1(\omega_{f g}(\alpha,Q^2))\big], \label{avwtiP}
\end{equation}
where ($\hat \alpha = 1-\alpha$)
\begin{align}
%\omega(M_f^2,M_{\bar g}^2,\alpha,Q^2) &=& M_f^2 \hat \alpha + \alpha M_{\bar g}^2 + \alpha \hat\alpha Q^2\,,
\omega_{f g}(\alpha,Q^2) &= M_f^2 \hat \alpha + \alpha M_{g}^2 + \alpha \hat\alpha Q^2\,.
\label{eq:omega}
\end{align}

Using Eq.\,\eqref{avwtiP}, one arrives at the following Bethe-Salpeter equation:
\begin{equation}
\label{bsefinalE}
\left[
\begin{array}{c}
E_{0^-}(Q)\\
F_{0^-}(Q)
\end{array}
\right]
= \frac{4 \alpha_{\rm IR}}{3\pi m_G^2}
\left[
\begin{array}{cc}
{\cal K}_{EE}^{0^-} & {\cal K}_{EF}^{0^-} \\
{\cal K}_{FE}^{0^-} & {\cal K}_{FF}^{0^-}
\end{array}\right]
\left[\begin{array}{c}
E_{0^-}(Q)\\
F_{0^-}(Q)
\end{array}
\right],
\end{equation}
with
{\allowdisplaybreaks
\begin{subequations}
\label{fgKernel}
\begin{eqnarray}
\nonumber
{\cal K}_{EE}^{0^-} &=&
\int_0^1d\alpha \bigg\{
{\cal C}_0^{\rm iu}(\omega_{f g}( \alpha, Q^2))  \\
&&+ \bigg[ M_f M_{g}-\alpha \hat\alpha Q^2 - \omega_{f g}( \alpha, Q^2)\bigg]\nonumber \\
&&
\quad \times
\overline{\cal C}^{\rm iu}_1(\omega_{f g}(\alpha, Q^2))\bigg\},\\
\nonumber
{\cal K}_{EF}^{0^-} &=& \frac{Q^2}{2 M_{f g}} \int_0^1d\alpha\, \bigg[\hat \alpha M_f+\alpha M_{g}\bigg]\\
&& \quad \times \overline{\cal C}^{\rm iu}_1(\omega_{f g}(\alpha, Q^2)),\\
{\cal K}_{FE}^{0^-} &=& \frac{2 M_{f g}^2}{Q^2} {\cal K}_{EF}^{0^-} ,\\
\nonumber
{\cal K}_{FF}^{0^-} &=& - \frac{1}{2} \int_0^1d\alpha\, \bigg[ M_f M_{g}+\hat\alpha M_f^2+\alpha M_{g}^2\bigg]\\
&& \quad \times \overline{\cal C}^{\rm iu}_1(\omega_{f g}(\alpha, Q^2))\,.
\end{eqnarray}
\end{subequations}}

The value of $Q^2=-m_{0^-}^2$ for which Eq.\,\eqref{bsefinalE} is satisfied supplies the bound-state mass and the associated solution vector is the meson's Bethe-Salpeter amplitude.  In the calculation of observables, the canonically normalised amplitude must be used, \emph{viz}.\ the amplitude obtained after rescaling such that
\begin{equation}
\label{normcan}
1=\left. \frac{d}{d Q^2}\Pi_{0^-}(Z,Q)\right|_{Z=Q},
\end{equation}
where, with the trace over spinor indices:
\begin{align}
\Pi_{0^-}(Z,Q) & = 6 {\rm tr}_{\rm D} \!\! \int\! \frac{d^4t}{(2\pi)^4} \nonumber \\
& \quad \times  \Gamma_{0^-}(-Z)
 S_f(t_+) \, \Gamma_{0^-}(Z)\, S_g(t)\,.
 \label{normcan2}
\end{align}

In terms of the canonically normalised Bethe-Salpe\-ter amplitude, the pseudoscalar meson's leptonic decay constant is
\begin{align}
f_{0^-} &= \frac{N_c}{4\pi^2}\frac{1}{ M_{f g}}\,
\big[ E_{0^-} {\cal K}_{FE}^{0^-} + F_{0^-}{\cal K}_{FF}^{0^-} \big]_{Q^2=-m_{0^-}^2}\,. \label{ffg}
\end{align}

\begin{table}[t]
\caption{\label{Tab:DressedQuarks}
Couplings, ultraviolet cutoffs and current-quark masses that deliver a good description of flavoured pseudoscalar meson properties, along with the dressed-quark masses and chosen pseudoscalar meson properties they produce; all obtained with $m_G=0.5\,$GeV, $\Lambda_{\rm ir} = 0.24\,$GeV.
Empirically, at a sensible level of precision \cite{Zyla:2020zbs}:
$m_\pi =0.14$, $f_\pi=0.092$;
$m_K=0.50$, $f_K=0.11$;
$m_{D} =1.87$, $f_{D}=0.15$;
$m_{B}=5.30$, $f_{B}=0.14$.
%A value of $f_{\eta_b}=0.47$ is taken from lQCD \cite{McNeile:2012qf}.
%
(Dimensioned quantities in GeV.)}
\begin{center}
\begin{tabular*}%{|c|c|c|c|c|c|c|}\hline
{\hsize}
{
l@{\extracolsep{0ptplus1fil}}|
c@{\extracolsep{0ptplus1fil}}|
c@{\extracolsep{0ptplus1fil}}
c@{\extracolsep{0ptplus1fil}}
c@{\extracolsep{0ptplus1fil}}
|c@{\extracolsep{0ptplus1fil}}
c@{\extracolsep{0ptplus1fil}}
c@{\extracolsep{0ptplus1fil}}}\hline\hline
& quark & $\alpha_{\rm IR}/\pi\ $ & $\Lambda_{\rm uv}$ & $m$ &   $M$ &  $m_{0^-}$ & $f_{0^-}$ \\\hline
%$l=u/d\ $  & $0.93$ & 0.905 & 0.007 & 0.367 & 0.14 & 0.101  \\\hline
%$\pi\ $  & $l=u/d\ $  & $0.93$ & $0.91\ $ & $0.007\ $ & 0.37$\ $ & 0.14 & 0.10  \\\hline
$\pi\ $  & $l=u/d\ $  & $0.36\phantom{2}$ & $0.91\ $ & $0.007\ $ & 0.37$\ $ & 0.14 & 0.10  \\\hline
%$s$  & $0.93 $ & 0.905 & 0.17 & 0.533 & 0.50 & 0.106 \\\hline
%$K\ $ & $\bar s$  & $0.84$ & $0.94\ $ & $0.16\phantom{7}\ $ & 0.53$\ $ & 0.50 & 0.11 \\\hline
$K\ $ & $\bar s$  & $0.33\phantom{2}$ & $0.94\ $ & $0.16\phantom{7}\ $ & 0.53$\ $ & 0.50 & 0.11 \\\hline
%$c$  & 0.438 & 1.878 & 1.235 & 1.603 & 3.177 & \\\hline
%$D\ $ & $c$  & $0.32$ & $1.36\ $ & $1.39\phantom{7}\ $ & 1.57$\ $ & 1.87 & 0.15 \\\hline
$D\ $ & $c$  & $0.12\phantom{2}$ & $1.36\ $ & $1.39\phantom{7}\ $ & 1.57$\ $ & 1.87 & 0.15 \\\hline%$b$  & 0.097 & 3.495 & 4.669 & 4.829 & 3.175 &
%$B\ $ & $\bar b$  & $0.13$ & $1.92\ $ & $4.81\phantom{7}\ $ & 4.81$\ $ & 5.30 & 0.14 \\
$B\ $ & $\bar b$  & $0.052$ & $1.92\ $ & $4.81\phantom{7}\ $ & 4.81$\ $ & 5.30 & 0.14
\\\hline\hline
% bbar 0.0116506 Pi
\end{tabular*}
\end{center}
\end{table}

%% Jenny's scheme
%% Mq & Lambda_uv are varied in order to obtain given values of m_meson & f_meson
The properties of $\pi$- and $\rho$-mesons were analysed in Ref.\,\cite{Roberts:2011wy}, with the optimal description provided by the parameters and associated current-quark mass in the middle columns of Table~\ref{Tab:DressedQuarks}, row~1.  The last three columns report calculated results for the dressed $u$-quark mass, pion mass and pion decay constant.

Keeping the light-quark values, we determined the $s$-quark current mass, $m_s$, and $K$-meson ultraviolet cutoff, $\Lambda_{\rm uv}^K$, through a least-squares fit to empirical values of $m_K$, $f_K$ whilst imposing the relation:
\begin{equation}
\alpha_{\rm IR}(\Lambda_{\rm uv}^{K}) [\Lambda_{\rm uv}^{K}]^2 \ln\frac{\Lambda_{\rm uv}^{K}}{\Lambda_{\rm ir}}
=
\alpha_{\rm IR}(\Lambda_{\rm uv}^{\pi}) [\Lambda_{\rm uv}^{\pi}]^2 \ln\frac{\Lambda_{\rm uv}^{\pi}}{\Lambda_{\rm ir}}\,.
\label{alphaLambda}
\end{equation}
This procedure eliminates one parameter by implementing the physical constraint that any increase in the momentum-space extent of a hadron wave function is matched by a reduction in the effective coupling between the constituents. Critical over-binding is thus avoided.  The results are listed in Table~\ref{Tab:DressedQuarks}, row~2.
The procedure is repeated for the $c$-quark/$D$-meson and $\bar b$-quark/$B$-meson, with results in Table~\ref{Tab:DressedQuarks}, rows~3, 4.

Regarding the current-quark masses in Table~\ref{Tab:DressedQuarks}, the fitted value of $m_s/m_l = 24$ matches estimates in QCD \cite{Zyla:2020zbs}, despite the individual current-masses being too large by a factor of $\lesssim 2$ because the SCI is deficient in connection with ultraviolet quantities.
The result $M_s/M_l =1.4$ is a fair match with the value determined in efficacious RL studies with momentum-dependent interactions \cite{Roberts:2020udq}: $M_s/M_l =1.25(9)$.
Similarly, the results for $m_c$, $m_b$ are individually somewhat higher than QCD estimates, but the values for $M_{c,b}$ are commensurate with typical values of the heavy-quark pole masses \cite{Zyla:2020zbs}.
%  1.275 ... mc/ms = 11.7 ... pole mass = 1.67(7)
% 4.18 ... mb/mc = 4.53 ... pole mass = 4.78(6)
% mb-mc = 3.45(5)
This is the character of the SCI: it is not a precision tool; but when employed judiciously, it is qualitatively and semiquantitatively reliable.

\begin{table}[t]
\caption{\label{Tab:MesonSpectrum}
%%%
Computed masses, Bethe-Salpeter amplitudes, and decay constants for a representative selection of mesons.
Empirical masses from Ref.\,\cite{Zyla:2020zbs};
%$m_{B^\ast_c}$, $m_{B_{c}^{0^{+}}}$, $m_{B_{c}^{1^{+}}}$,
entry marked by ``$\ast$'' from Ref.\,\cite{Mathur:2018epb}.
Empirically unknown decay constants quoted from lQCD \cite{Becirevic:1998ua, Chiu:2007bc, Colquhoun:2015oha, Aoki:2019cca}.
%
%\cite{Aoki:2019cca}
%% Aoki:2019cca
%% Becirevic:1998ua ... quenched
%%% Davies:2010ip f_etac ... but not the same as ours
%% Donald:2012ga J/Psi
%
(Dimensioned quantities in GeV.
Underlined entries from Table~\ref{Tab:DressedQuarks}.)
%Underlined entries were used to fit current-quark masses and CI parameters.
%Using the CI, only $0^-$ mesons possess a $F$ amplitude; so, these sites are otherwise left empty.  `
%``---'' indicates no empirical/lQCD results available for comparison.)
}
%%%
%\begin{center}
\begin{tabular*}%{|c|c|c|c|c|c|c|}\hline
{\hsize}
{c@{\extracolsep{0ptplus1fil}}
l@{\extracolsep{0ptplus1fil}}||
l@{\extracolsep{0ptplus1fil}}
l@{\extracolsep{0ptplus1fil}}|
l@{\extracolsep{0ptplus1fil}}
l@{\extracolsep{0ptplus1fil}}|
l@{\extracolsep{0ptplus1fil}}
l@{\extracolsep{0ptplus1fil}}
}
\hline\hline
$J^{P}$ &      Meson               &     $m^{\rm CI}$     &        $m^{\rm e/l}$         &         $E$      &      $F$         &       $f^{\rm CI}$         &       $f^{\rm e/l}$   \\
\hline
$0^{-}$ & $\pi(u\bar{d})$          &      \underline{0.14}       &          0.14            &        3.59     &     0.47$\ $        &         \underline{0.10}          &           0.092         \\
        & $K(u\bar{s})$            &      \underline{0.50}       &          0.50            &        3.70     &     0.55$\ $        &         \underline{0.11}          &           0.11         \\

        & $D(u\bar{c})$            &     \underline{1.87}       &  1.87    &        3.25     &     0.39$\ $        &         0.15          &           0.15(1)         \\
        & $D_{s}(s\bar{c})$        &      1.96       &          1.97            &        3.45     &     0.54$\ $        &         0.16          &           0.18         \\
        & $\eta_{c}(c\bar{c})$     &      2.90       &          2.98            &        3.74     &     0.90$\ $        &         0.20 &           0.24(1)         \\  % expt
        & $B(u\bar{b})$            &      \underline{5.30}    &          5.30            &        2.98     &     0.18$\ $        &         \underline{0.14}      &           0.13         \\
        & $B_{s}(s\bar{b})$        &      5.38       &          5.37            &        3.26     &     0.27$\ $        &         0.16          &           0.16         \\
        & $B_{c}(c\bar{b})$        &      6.16       &          6.28            &        4.25     &     0.79$\ $        &         0.21          &           0.35         \\
\hline
%$1^{-}$ & $\rho(u\bar{d})$         &      0.93       &          0.78            &        1.53     &              &         0.13          &           0.15         \\
%
$1^{-}$   & $K^{\ast}(u\bar{s})$     &      1.10       &          0.89            &        1.31     &               &         0.15          &           0.16         \\
        & $D^{\ast}(u\bar{c})$     &      2.09       &          2.01            &        1.25     &               &         0.15          &           0.17(1)         \\
        & $D_{s}^{\ast}(s\bar{c})$ &      2.18       &          2.11            &        1.30     &              &         0.15          &           0.19(1)         \\
        & $B^{\ast}(u\bar{b})$     &      5.36       &          5.33            &        1.26     &               &         0.13          &           0.13(2)         \\
        & $B_{s}^{\ast}(s\bar{b})$ &      5.45       &          5.42            &        1.34     &               &         0.14          &           0.16(2)         \\
        & $B_{c}^{\ast}(c\bar{b})$ &      6.24       &          6.33$^\ast$     &        1.97     &              &         0.20          &           0.34(2)         \\
\hline\hline
\end{tabular*}
%\end{center}
\end{table}

The evolution of the ultraviolet cutoff with pseudoscalar meson mass reported in Table~\ref{Tab:DressedQuarks} is described by the following interpolation $(s=m_{0^-}^2)$:
\begin{equation}
\label{LambdaIRMass}
\Lambda_{\rm uv}(s) = 0.306 \ln [ 19.2 + (s/m_\pi^2-1)/2.70]\,.
\end{equation}
For a given meson, $H$, the associated coupling can then be obtained using Eq.\,\eqref{alphaLambda}, with $[\Lambda_{\rm uv}^{K}]^2 \to m_H^2=s$.  One can subsequently compute properties of any meson for which a mass estimate is available by solving the associated Bethe-Salpeter equation with the prescribed cutoff and coupling, using the dressed-quark propagators already determined.
%  Gap equation solution is always available now.  Only need value of cutoff in BSE.  That is fixed by empirical mass.
Proceeding as described here, one obtains the meson masses and decay constants listed in Table~\ref{Tab:MesonSpectrum}.

\begin{figure}[t]
\vspace*{2ex}

\leftline{\hspace*{0.5em}{\large{\textsf{A}}}}
\vspace*{-3ex}
\includegraphics[width=0.45\textwidth]{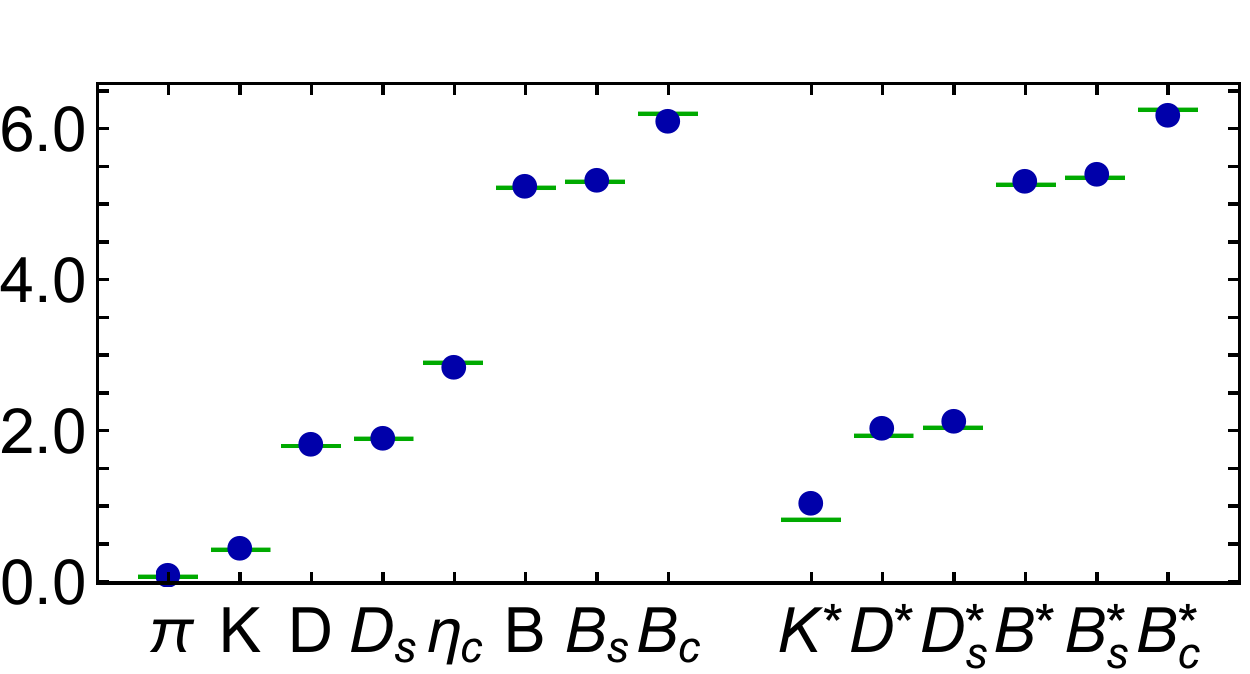}
\vspace*{1ex}

\leftline{\hspace*{0.5em}{\large{\textsf{B}}}}
\vspace*{-3ex}
\includegraphics[width=0.45\textwidth]{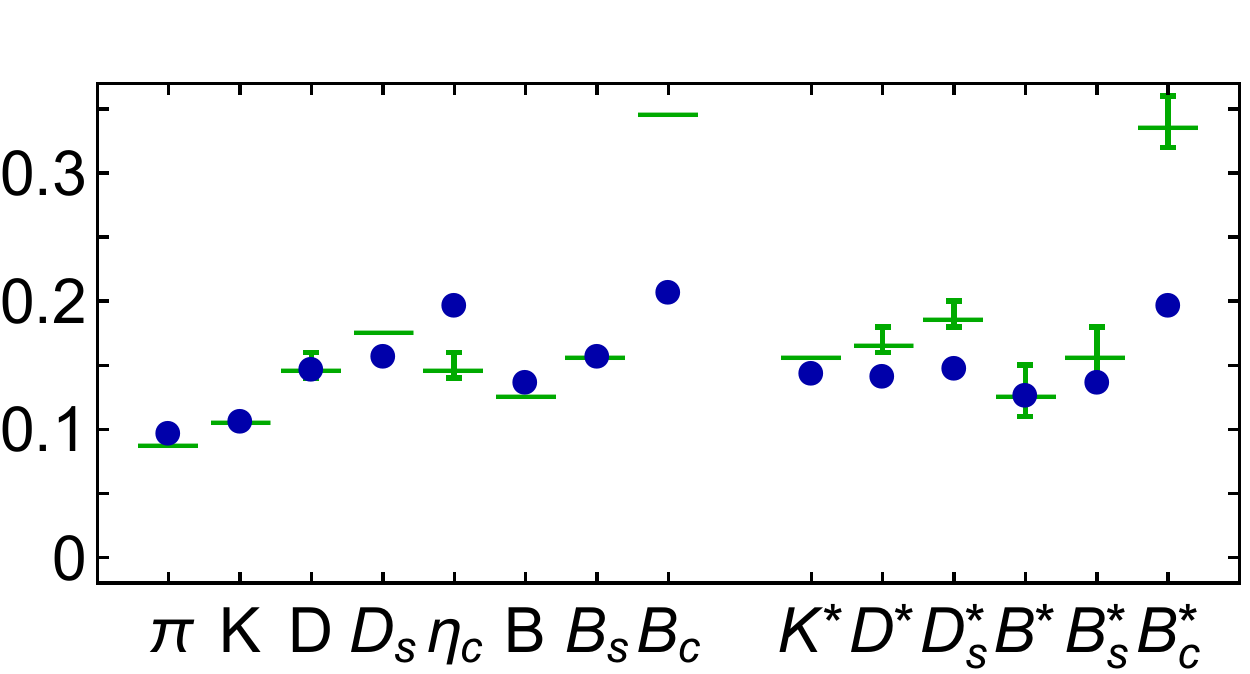}

\caption{\label{FigMeson}
\emph{Upper panel}\,--\,{\sf A}.
Comparison between CI predictions for selected meson masses and available experiment \cite{Zyla:2020zbs} and $m_{B^\ast_c}$ from Ref.\,\cite{Mathur:2018epb}.
\emph{Lower panel}\,--\,{\sf B}.
Analogous comparison for meson leptonic decay constants: experiment \cite{Zyla:2020zbs}, where known, and lQCD otherwise \cite{Becirevic:1998ua, Chiu:2007bc, Colquhoun:2015oha, Aoki:2019cca}.
In both panels, contact-interaction predictions are depicted as blue circles and comparison values by green bars.
(Pictorial representation of results in Table~\ref{Tab:MesonSpectrum}.)
}
\end{figure}

As highlighted in Fig.\,\ref{FigMeson}A, SCI results for the masses are in good agreement with experiment: the mean absolute relative difference $\overline{\rm ard} =2.8$\%.  Particularities associated with ensuring the vector Ward-Green-Takahashi identity entail that the description is better for pseudoscalars than it is for vector mesons \cite{Roberts:2010rn}.

Regarding the leptonic decay constants, Fig.\,\ref{FigMeson}B, the SCI supplies a poorer description because these observables describe quark+antiquark annihilation at a single spacetime point.  Hence, they are sensitive to ultraviolet physics, which is a challenge for the SCI.  Eqs.\,\eqref{alphaLambda}, \eqref{LambdaIRMass} are useful in repairing the deficiency.  They ensure that in comparison with known empirical values or available lQCD results, the picture is fair: trends are typically reproduced; and comparing columns~6 and 7 in Table~\ref{Tab:MesonSpectrum}, $\overline{\rm ard} =13$\%.
The discrepancies between SCI and lQCD results for $f_{B_c}$, $f_{B_c^\ast}$ appear anomalously large.  Eliminating them, then $\overline{\rm ard} =8.3$\%.  However, although the lQCD results were obtained some time ago \cite{Chiu:2007bc}, they are consistent with modern continuum predictions developed using realistic interactions \cite{Binosi:2018rht}.

It should be noted that our ultraviolet improvement scheme differs from that in Ref.\,\cite{Yin:2021uom}.  Therein, pseudoscalar mesons with mass-degenerate valence degrees-of-freedom were used to develop a running scale like that in Eq.\,\eqref{LambdaIRMass}.  Owing to our subsequent focus on semileptonic decays of flavoured pseudoscalar mesons, we instead used the underlined masses in Table~\ref{Tab:MesonSpectrum}.
%
%Further improvement can be achieved by additional tuning of Eq.\,\eqref{alphaLambda}, including its extension into the light-quark sector and consideration of the particular properties of heavy+light mesons.

\section{Semileptonic Transitions: Foundations}
\label{SecSemiLep}
\subsection{Matrix Elements}
Our SCI analysis of pseudoscalar meson semileptonic decays follows the treatment of $K_{\ell 3}$ transitions in Ref.\,\cite{Chen:2012txa}.  We will sketch the procedure using the $D^0\to \pi^-$ transition.  All others can be developed by analogy.  Therefore, consider the following matrix element:\footnote{Euclidean metric conventions:  $\{\gamma_\mu,\gamma_\nu\} = 2\delta_{\mu\nu}$; $\gamma_\mu^\dagger = \gamma_\mu$; $\gamma_5= \gamma_4\gamma_1\gamma_2\gamma_3$, tr$[\gamma_5\gamma_\mu\gamma_\nu\gamma_\rho\gamma_\sigma]=-4 \epsilon_{\mu\nu\rho\sigma}$; $\sigma_{\mu\nu}=(i/2)[\gamma_\mu,\gamma_\nu]$; $a \cdot b = \sum_{i=1}^4 a_i b_i$; and $Q_\mu$ timelike $\Rightarrow$ $Q^2<0$.}
\begin{align}
_d  M_\mu^{D^0}(P,Q) &=\langle \pi^-(p) | \bar d i\gamma_\mu c |D^0(k)\rangle  \nonumber \\
& = [P_\mu f_+^{D_u^d}(t) + Q_\mu f_-^{D_u^d}(t)]\,,
\label{TypicalME}
\end{align}
where
$P =k+p$,
$Q=p-k$,
with $k^2 = -m_{D}^2$ and $p^2=-m_{\pi}^2$;
and the squared-momentum-transfer is $t=-Q^2$.

The masses of the hadrons involved limit the physical domain of support for the form factors:
\begin{subequations}
\begin{align}
P\cdot Q & = - (m_{D}^2 - m_{\pi}^2) =: - \Delta_{D \pi}\,,\\
P^2 & = - 2 (m_{D}^2 + m_{\pi}^2) - Q^2 =: -2 \Sigma_{D\pi} - Q^2;
\end{align}
\end{subequations}
and $t_m^{D\pi} = (m_{D} - m_{\pi})^2 =: m_{D}^2 y_m^{D\pi} $  is the largest value of the squared-momentum-transfer in the decay process.

In the flavour-symmetry limit, $f_+^{D}(t)$ is equivalent to the elastic form factor of a charged pion-like meson built from a valence-quark and -antiquark with degenerate current masses and $f_-^{D}(t)\equiv 0$ \cite{Chen:2018rwz}.  In all transitions, therefore, $f_-(t)$ is sensitive to the strength of HB-induced flavour-symmetry breaking, as are various sensibly constructed ratios of $f_+$ transition form factors, \emph{e.g}.\ $D^+ \to K^0$ \emph{vs}.\ $D^0 \to \pi^-$ reflects $s$-quark:$u$-quark differences.
These properties are correlated with the scalar form factor
\begin{equation}
\label{Eqf0}
f_0^D(t) = f_+^D(t) + \frac{t}{m_{D}^2 - m_{\pi}^2} f_-^D(t)\,,
\end{equation}
and its analogues, which all measure the divergence of the transition current, $Q\cdot M(P,Q)$.

The additional merit of focusing on $f_{+,0}(t)$ is that each is characterised by a different resonance structure on $t \gtrsim t_m$: $f_{+}(t)$ links with the vector meson $D^\ast$; and $f_{0}(t)$ with the related scalar resonance.  On the other hand, $f_{-}(t)$ overlaps with both channels.  (These observations are exemplified in Refs.\,\cite{Ji:2001pj, Chen:2012txa, Yao:2020vef} and below.)

Using our SCI, the matrix element in Eq.\,\eqref{TypicalME} takes the following explicit form:
\begin{align}
_d  M_\mu^{D^0}& (P,Q) =
N_c {\rm tr}\int\frac{d^4 t}{(2\pi)^4}
\Gamma_{D}(p) S_c(t+p) \nonumber \\
& \times i \Gamma_\mu^{cd}(Q) S_d(t+k) \Gamma_\pi(-k) S_u(t)\,,
\label{dMD}
\end{align}
where the trace is over spinor indices and $N_c=3$.  The scalar functions characterising the transition are obtained from Eq.\,\eqref{dMD} using straightforward projections:
\begin{subequations}
\label{Projections}
\begin{align}
  f_+^{D_u^d}(t) & = \frac{t P_\mu - (m_{D}^2- m_\pi^2) Q_\mu}{t P^2 + (m_{D}^2- m_\pi^2)^2}
\, _d M_\mu^{D^0}(P,Q)\,, \\
 f_0^{D_u^d}(t) & = - \frac{Q_\mu}{m_{D}^2- m_\pi^2} \, _d M_\mu^{D^0}(P,Q)\,,
\end{align}
\end{subequations}
with $f_-^{D_u^d}(t)$ reconstructed via Eq.\,\eqref{Eqf0}.

Three types of matrix-valued functions appear in Eq.\,\eqref{dMD}.
Simplest are the dressed-quark propagators, which have the form in Eq.\,\eqref{genS} and involve the appropriate dressed-quark mass drawn from Table~\ref{Tab:DressedQuarks}.
Next are the Bethe-Salpeter amplitudes for the participating mesons, whose structure is given in Eq.\,\eqref{PSBSA} and which are also completed by the numerical results in Table~\ref{Tab:DressedQuarks}.
The new element is the dressed vector component of, in this instance, the $c\to d$ weak transition vertex, $\Gamma_\mu^{cd}$.  Other transitions involve an analogous vector vertex.

\subsection{Weak Vector Vertex}
\label{WeakVectorVertex}
The vector component of the $c\to d$ weak transition vertex satisfies a Ward-Green-Takahashi identity:
\begin{equation}
\label{WGTI}
Q_{\mu} i \Gamma^{cd}_{\mu}(Q) = S_c^{-1}(t+Q) - S_d^{-1}(t)- (m_c-m_d)\Gamma^{cd}_I(Q)\,,
\end{equation}
where $\Gamma_I^{cu}$ is an analogous Dirac-scalar vertex.  Recall that the axial-vector piece of the quark weak vertex does not contribute to a $0^- \to 0^-$ transition.  As with all $n$-point functions, care must be taken when formulating the SCI solution procedure for this vertex; so, it is worth providing some details.

{\allowdisplaybreaks
The two vertices in Eq.\,\eqref{WGTI} satisfy inhomogeneous Bethe-Salpeter equations, \emph{viz}.\ in RL truncation:
\begin{subequations}
\begin{align}
\Gamma_\mu^{cd} (Q) & = \gamma_\mu  - \frac{16}{3} \frac{\pi \alpha_{\rm IR}}{m_G^2} \nonumber \\
& \times
 \int\frac{d^4 t}{(2\pi)^4} \! \gamma_\alpha S_c(t+Q) \Gamma_\mu^{cd}(Q) S_d(t)\gamma_\alpha\,,
 \label{vectorIBSE}  \\
\nonumber
\Gamma^{cu}_I(Q) & = I_{\rm D} - \frac{16}{3} \frac{\pi \alpha_{\rm IR}}{m_G^2} \\
& \times  \int\! \frac{d^4t}{(2\pi)^4}
\gamma_\alpha S_c(t+Q) \Gamma^{su}_I(Q) S_d(t) \gamma_\alpha\,. \label{scalarIBSE}
\end{align}
\end{subequations}
So long as the regularisation scheme is symmetry preserving, then the solutions are
\begin{subequations}
\begin{align}
\Gamma^{cd}_\mu(Q) & = \gamma_\mu^T P_T^{cd}(Q^2) \nonumber \\
& \qquad + \gamma_\mu^L P_{1L}^{cd}(Q^2) - i Q_\mu I_{\rm D} P_{2L}^{cd}(Q^2)\,,
\label{genvector}\\
\Gamma^{cd}_I(Q) & = I_{\rm D} \, {E}_I^{cd}(Q^2) \,,
\label{genscalar}
\end{align}
\end{subequations}
where $I_{\rm D}$ is the identity matrix in spinor space, $Q_\mu \gamma_\mu^T = 0$, $\gamma_\mu^T+\gamma_\mu^L = \gamma_\mu$.

Following Ref.\,\cite{Chen:2012txa}, one finds:
\begin{subequations}
\begin{equation}
\label{answerScalar}
{E}_I^{cd}(Q^2) = \frac{1}{1+K_{E}^{cd}(Q^2)}\,,
\end{equation}
where
\begin{align}
\nonumber
K_{E}^{cd}(Q^2) &=  -\frac{4\alpha_{\rm IR}}{3\pi m_G^2} \int_0^1 d\alpha \bigg[
{\cal C}^{\rm iu}(w_{d\bar c})
- {\cal C}^{\rm iu}_1(w_{d\bar c})\\
& \qquad - (M_d M_c + \alpha \hat \alpha Q^2) \, \overline {\cal C}^{\rm iu}_1(w_{d\bar c})\bigg],
\label{KEsu}
\end{align}
\end{subequations}
$w_{d\bar c}=\omega_{d\bar c}(\alpha,Q^2)$, Eq.\,\eqref{eq:omega}, and the other functions are defined via Eq.\,\eqref{eq:Cn}.
It is apparent under further inspection that ${E}_I^{cd}(Q^2)$ exhibits a pole at the mass of the lightest $0^+$ $\bar c d$ state, \emph{i.e}.\ the $D_0^{\ast}$-meson.  Furthermore,
\begin{equation}
\label{mdmdE}
 (m_c-m_d) {E}_I^{cd}(Q^2=0) = M_c - M_d \,.
\end{equation}
}

It is worth remarking that owing to uncertainties in the spectrum of heavy+light $0^+$ mesons, the coupling and ultraviolet cutoff in Eq.\,\eqref{KEsu} are fixed using Eqs.\,\eqref{alphaLambda}, \eqref{LambdaIRMass} evaluated at the mass of the ground-state pseudoscalar meson in the given channel; here, $m_D$.
%% The c-quark current mass, mc, is tuned so that with this same cutoff, Mc = fixed.
As noted above, ultraviolet cutoffs play a dynamical role in the formulation of a contact interaction; so, this choice is merely part of the definition of the symmetry-preserving regularisation scheme.  Given that it is a scalar channel, the value of the ultraviolet cutoff feeds back through Eq.\,\eqref{mdmdE} into small variations of the current-quark masses so that the dressed-quark masses remain fixed.

{\allowdisplaybreaks
Turning to Eq.\,\eqref{vectorIBSE}, one readily finds \cite{Chen:2012txa}
\begin{equation}
P_{1L}^{cd}(Q^2) \equiv 1\,;
\end{equation}
and
\begin{align}
\label{PTsu}
P_T^{cd}(Q^2) &=  \frac{1}{1+K_V^{cd}(Q^2)}\,,\\
\nonumber
K_V^{cd}(Q^2)&= -\frac{2\alpha_{\rm IR}}{3\pi m_G^2} \int_0^1 d\alpha \big[
M_d M_c - M_d^2 \hat\alpha \\
& \qquad - M_c^2\alpha-2 \alpha\hat\alpha Q^2\big]
\overline {\cal C}^{\rm iu}_1(w_{d\bar c})\,.
\end{align}
The transverse part of this quark weak vector vertex has a pole at the mass of the $D^\ast$-meson.
}

Considering Eq.\,\eqref{genvector}, only $P_{2L}^{cd}(Q^2)$ remains undetermined.  Adapting the analysis in Ref.\,\cite{Chen:2012txa}, one obtains
\begin{equation}
%Q^2 P^{su}_{2L}(Q^2) = (m_u-m_s) {\cal E}_I^{su}(Q^2) - M_u+M_s\,.
Q^2 P^{cd}_{2L}(Q^2) = (m_d-m_c) {E}_I^{cd}(Q^2) - M_d+M_c\,.
\end{equation}
Using Eq.\,\eqref{mdmdE}, it is clear that $\lim_{Q^2\to 0} Q^2 P^{cd}_{2L}(Q^2)=0$; consequently, $\Gamma^{cd}_\mu(Q)$ is regular at $Q^2=0$.

Our attempt to achieve a symmetry-preserving calculation of the quark weak vector vertex has almost succeeded.   One issue remains; namely, consistency between Eqs.\,\eqref{genS}, \eqref{WGTI}, \eqref{mdmdE} requires
\begin{equation}
P_T^{cd}(Q^2=0) = 1\,.
\end{equation}
Considering the ten transitions considered herein, our scheme fails by $2.4(1.5)$\%.  We subsequently correct for this by modifying the denominator in Eq.\,\eqref{PTsu}:
\begin{equation}
K_V^{cd}(Q^2) \to \hat K_V^{cd}(Q^2) := K_V^{cd}(Q^2)-K_V^{cd}(0)\,,
\end{equation}
with analogous corrections in the other vertices.  This is akin to implementing momentum-subtraction regularisations of the vector-boson vacuum polarisations.

\begin{table*}[t]
\caption{\label{fp0val}
\emph{Upper panel}\,--\,{\sf A}.
Maximum recoil $(t=0)$ value of semileptonic transition form factors compared with (where available) inferences from experiment \cite{Ablikim:2015ixa, Ablikim:2018upe}
%
%and lQCD \cite{Bouchard:2014ypa, Lubicz:2017syv} (where available).
and lQCD \cite{Aoki:2019cca, McLean:2019qcx, Colquhoun:2016osw, Cooper:2020wnj}.
In each case, we list results for $f_+^{H_1 \to H_2}(0)$ obtained with (column 1) the complete SCI Bethe-Salpeter amplitude, Eq.\,\eqref{PSBSA}, and (column 2) with $F_{0^-}\to 0$ and all procedures repeated.  Comparison between these columns provides an indication of the sensitivity in SCI predictions to formulation details.
Our analogous predictions for $f_-^{H_1 \to H_2}(t=0)$ are listed in columns 6 and 7.
\emph{Lower panel}\,--\,{\sf B}.
SCI computed branching fractions (columns~1, 2), obtained using empirical masses, compared with, where available:
continuum results from Ref.\,\cite{Yao:2020vef} (column~3);
empirical results (column~4) \cite{Ablikim:2015ixa, Ablikim:2018upe, Sibidanov:2013rkk, Zyla:2020zbs};
lQCD (column 5) \cite{Cooper:2020wnj};
and illustrative model analyses (column~6) \cite{Ebert:2003wc, Zhao:2006at, Faustov:2014bxa, Hu:2019qcn}.
(All numerical entries should be multiplied by $10^{-3}$.)
Ref.\,\cite{Zyla:2020zbs} lists
$|V_{cd}| = 0.221(4)$, $|V_{cs}|= 0.987(11)$
$|V_{ub}| = 0.00382(24)$, $|V_{cb}|= 0.0410(14)$; and the following lifetimes (in seconds):
$\tau_{D^0} = 4.10 \times 10^{-13}$,
$\tau_{D_s^\pm} = 5.04 \times 10^{-13}$,
$\tau_{B^0} = 1.519 \times 10^{-12}$,
$\tau_{B_s^0} = 1.515 \times 10^{-12}$,
$\tau_{B_c^\pm} = 5.10 \times 10^{-13}$.
}
\begin{center}
\begin{tabular}{l|cc|ccc||ccc}\hline
 {\sf A.}\ $f_+(0)$ & here$_{F_{0^-}\neq 0}$ & here$_{F_{0^-} \equiv 0}$ & Ref.\,\cite{Yao:2020vef} & expt. & lQCD & here$_{F_{0^-}\neq 0}^{|f_-(0)|}$ & here$_{F_{0^-}\equiv 0}^{|f_-(0)|}$ & Ref.\,\cite{Yao:2020vef}$_{|f_-(0)|}$\\\hline
% ... Vcd
\;\,1 \ $D\to \pi$ & $0.76$ & $0.65$ & $0.618(31)$ & $0.637(09)$ & $0.612(35)$  & $0.35$ & $0.54$ & 0.362(28)\\
\;\,2 \ $D_s\to K$ & $0.75$ & $0.65$ & $0.673(40)$ & $0.720(85)$  &   &$0.35$  & $0.55$ & 0.553(65)\\
% ... Vcs
\;\,3 \ $D\to K$ & $0.81$ & $0.72$ & $0.756(36)$ & $0.737(04)$ & $0.765(31)$ & $0.33$ &$0.50$ &  $0.277(45)$  \\
% ... Vub
\;\,4 \ $B\to \pi$ & $0.57$ & $0.48$ &  &  & $0.169(51)$ &  $0.38$  & $0.48$ & \\
\;\,5 \ $B_s\to K$ & $0.53$ & $0.43$ & &  & $0.281(12)$ & $0.36$  & $0.45$ & \\
% ... Vcb
\;\,6 \ $B \to D$ & $0.75$ & $0.72$ & &  & $0.818(22)$ & $0.40$  & $0.45$ & \\
\;\,7 \ $B_s \to D_s$ & $0.74$ & $0.71$ & &  & $0.666(12)$ & $0.39$  & $0.45$ & \\
\;\,8 \ $B_c \to \eta_c$ & $0.73$ & $0.61$ & &  & $\approx 0.6$  & $0.40$  & $0.43$ & \\
% ... Vub
\;\,9 \ $B_c \to B$ & $0.68$ & $0.67$  & &  & $0.55(2)$  & $0.49$  & $0.71$ & \\
% ... Vcs
10 \ $B_c \to B_s$ & $0.75$ & $0.75$ & &  &  $0.62(1)$ & $0.55$  & $0.79$ & \\
\hline
\end{tabular}\vspace*{1em}

\begin{tabular}{rl|cc|cccc}\hline
{\sf B.}\ & & here$_{F_{0^-}\neq 0}$ & here$_{F_{0^-} \equiv 0}$ & Ref.\,\cite{Yao:2020vef} & expt. & lQCD & model \\\hline
1 & ${\mathpzc B}_{D^0 \to \pi^- e^+ \nu_e}$ & $\ 2.96$ & $\ 2.57$ & $2.73(22)$ & $\ 2.95(05)$ & & \\
2 & ${\mathpzc B}_{D_s^+ \to K^0 e^+ \nu_e}$ & $\ 2.83$ & $\ 2.40$ & 3.31(33) & $\ 3.25(38)$ & & \\
3 & ${\mathpzc B}_{D^0 \to K^- e^+ \nu_e}$  & $37.0\phantom{0}$ & $31.8\phantom{0}$ & $38.3(2.8)$ & $35.05(36)$ & & \\
4 & ${\mathpzc B}_{\bar B^0 \to \pi^+ \ell^- \bar\nu_\ell}$ & $\ 0.41$ & $\ 0.34$ &  & $\quad 0.149(11)$ &  & \\
5 & ${\mathpzc B}_{B_s^0\to K^+ \ell^- \bar\nu_\ell}$ & $\ 0.36$ & $\ 0.29$ &  & & & $\quad 0.164(17)$ \\
6 & ${\mathpzc B}_{B^0\to D^- \ell^+ \nu_\ell}$ & $25.5\phantom{0} $ & $24.0\phantom{0} $ &  & 23.1(1.0) & &  \\
7 & ${\mathpzc B}_{B_s^0\to D_s^- \ell^+ \nu_\ell}$ & $25.0\phantom{0} $ & $23.5\phantom{0} $ &  & & & 24(5)\phantom{641} \\
8 & ${\mathpzc B}_{B_c^-\to \eta_c \ell^- \bar\nu_\ell}$ & $\phantom{1}8.99$ & $\phantom{1}6.65$ &  & & & $\;8.2(1.9)$ \\
9 & ${\mathpzc B}_{B_c^-\to B^0 \ell^- \nu_\ell}$ & $\phantom{1}1.21$ & $\phantom{1}1.20$ &  & %& $1.59(13)$  & $\ 1.4(1.0)$ \\
& $\phantom{1}0.87(5)$  & $\ 1.4(1.0)$ \\
10 &  ${\mathpzc B}_{B_c^+\to B_s^0 \ell^+\nu_\ell}$ & $17.9\phantom{0}$ & $18.1\phantom{9}$ %&  & & $26.9(1.4)\phantom{9}$ & $\ 16(11)\phantom{641}$ \\
&  & & $12.6(5)\phantom{9}$ & $\ 16(11)\phantom{641}$ \\
\hline
\end{tabular}
\end{center}
\end{table*}
%% D->pi FLAG  0.612(35) = Lubicz:2017syv
%% D->K FLAG = 0.765(31) = Lubicz:2017syv
%% B->Pi  FLAG ... my cm analysis 0.169(51)
%% Bs->K 0.323(63) 1406.2279 Bouchard:2014ypa
%% Bs->K  FLAG ... my cm analysis 0.238(37)
%% B->D  FLAG ... 0.818(22)
%% Bs->Ds McLean:2019qcx 0.666(12)
%% Bc->eta_c ... preliminary Colquhoun:2016osw
%% Bc-> B Cooper:2020wnj ... read from plots
%% Bc->Bs Cooper:2020wnj ... read from plots

% ... $B_c \to B$ links with $|V_{ud}|$;
% ... $B_c \to B_s$ with $|V_{us}|$;
% ... $B\to\pi$ and $B_s \to K$ with $|V_{ub}|$;
% ... $B\to D$, $B_s\to D_s$, $B_c \to \eta_c$ with $|V_{cb}|$.

%% ZNX
% For Bc->B:         up: 0.00086
%                     middle: 0.00081
%                        down: 0.00077
%For Bc->Bs:     up: 0.0131
%                  middle: 0.0126
%                    down: 0.0122

\subsection{Charge Conservation}
\label{sec:current}
In working from Eq.\,\eqref{dMD} to explicit expressions for \linebreak $f_+^{D_u^d}(t)$, $f_0^{D_u^d}(t)$, one follows the methods in Refs.\,\cite{Roberts:2011wy}.  They rely chiefly on the assumption of Poincar\'e-inva\-riance, including translational invariance in regularisation of the integrals.  Formally, the latter is always true.  However, the presence of the $\gamma_5\gamma\cdot Q$ term in Eq.\,\eqref{PSBSA} means it is broken in practice once Eq.\,\eqref{eq:C0} is used.  This violation is manifest in the elastic electromagnetic form factors of the pseudoscalar mesons that participate in the transitions, as we now explain.

Elastic form factors obtained using the SCI can be expressed as a sum of six terms.
The first separation stems from the photon coupling either to the quark ($\gamma q$) or the antiquark ($\gamma\bar q$).  It is the same for all quark+anti\-quark interactions.
When using the SCI, there are three subcomponents in each of these two terms.  Namely, the Bethe-Salpeter amplitude in Eq.\,\eqref{PSBSA} leads to $E_{0^-}^2$, $E_{0^-} F_{0^-}$, $F_{0^-}^2$ contributions.  The $F_{0^-}^2$ term vanishes at $Q^2=0$ and the $E_{0^-}^2$ is insensitive to the regularisation.  However, following the regularisation steps described hitherto
\begin{equation}
\label{EFneq}
\left. \langle E_{0^-} F_{0^-}\rangle \right|_{Q^2=0}^{\gamma q} \neq \langle \left. E_{0^-} F_{0^-}\rangle\right|_{Q^2=0}^{\gamma \bar q},
\end{equation}
\emph{i.e}.\ these two contributions are not equal and that violates charge conservation.  Typically, the relative error is $\lesssim 1$\%, but it should vanish.

The mismatch owes to the quadratic divergences that arise through integrals such as
\begin{eqnarray}
\nonumber
&&\int\frac{d^4 t}{(2\pi)^4} \frac{1}{[t^2+\omega]^2} \{(P\cdot t)^2,(Q\cdot t)^2, (P\cdot t) (Q\cdot t)\}\\
&=& \int\frac{d^4 t}{(2\pi)^4} \frac{t^2}{[t^2+\omega]^2}\frac{1}{4}
\{P^2,Q^2 , P\cdot Q \}\,, \label{eq:1on4}
\end{eqnarray}
which also affect the value of $f_+^{H_1 H_2}(0)$.  It can be remedied through a simple expedient: change
\begin{equation}
\frac{1}{4}  \to  \frac{1}{4}\,(1+\theta)
\end{equation}
and tune $\theta$ to restore equality of the two sides in Eq.\,\eqref{EFneq}.  We have computed $\theta$ from the elastic form factor of every one of the initial states in the transitions considered herein: $D_{(s)}$, $B_{(s,c})$, with the results $\theta = 0.28$, $0.40$, $0.01$, $0.18$, $0.17$.  (Evidently, the mismatch is eliminated by choosing $\theta \approx 0.21$.)  These values are used to complete the regularisation of the transition form factors specified by Eq.\,\eqref{dMD} and its analogues.
%Our analysis reproduces all results obtained in Ref.\,\cite{Chen:2012txa}.)

%
Before delivering new SCI predictions, it is worth remarking that we have confirmed all results in Ref.\,\cite{Chen:2012txa}, including those connected with $K_{\ell 3}$ and $\pi_{e3}$ transitions.

\begin{figure}[t]
\vspace*{2ex}

\leftline{\hspace*{0.5em}{\large{\textsf{A}}}}
\vspace*{-4ex}
\includegraphics[width=0.45\textwidth]{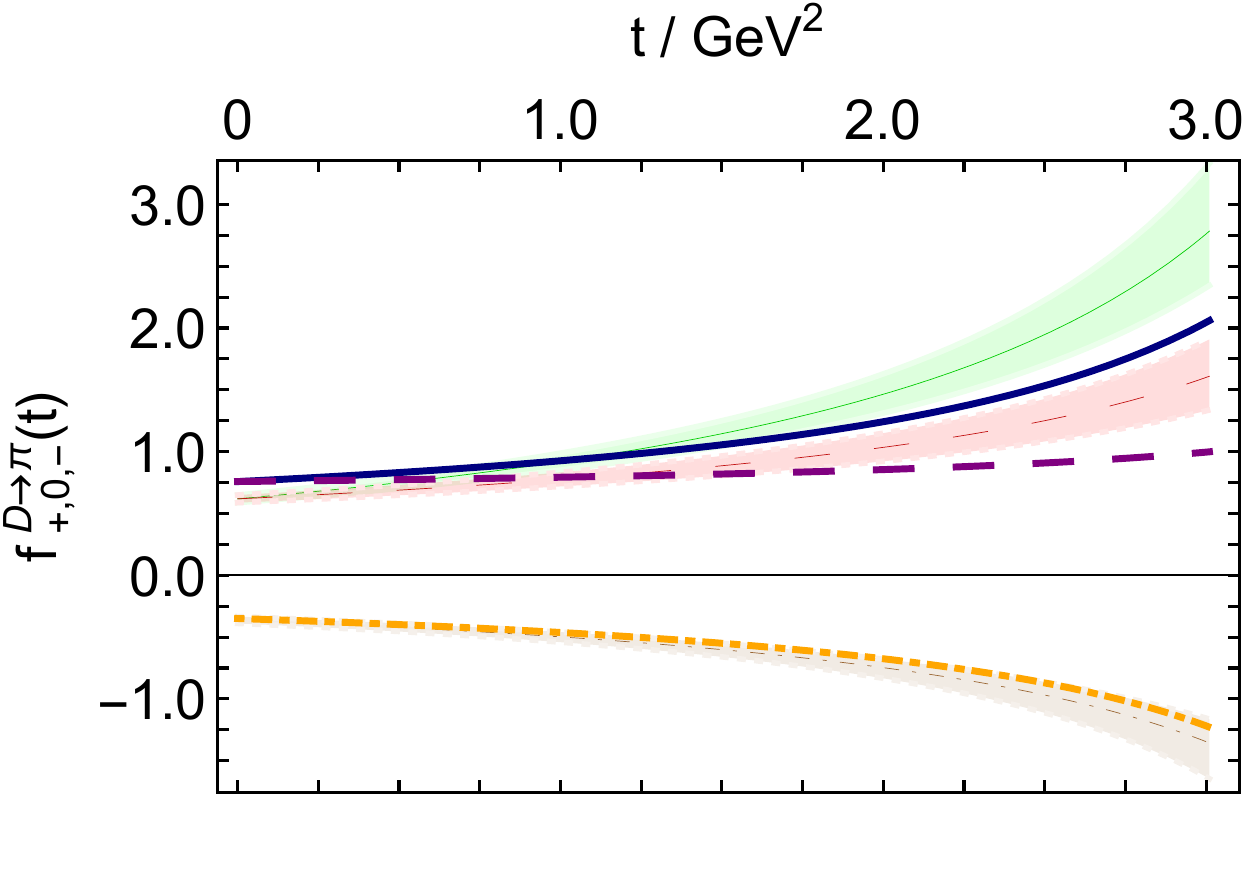}
\vspace*{-2ex}

\leftline{\hspace*{0.5em}{\large{\textsf{B}}}}
\vspace*{-4ex}
\includegraphics[width=0.45\textwidth]{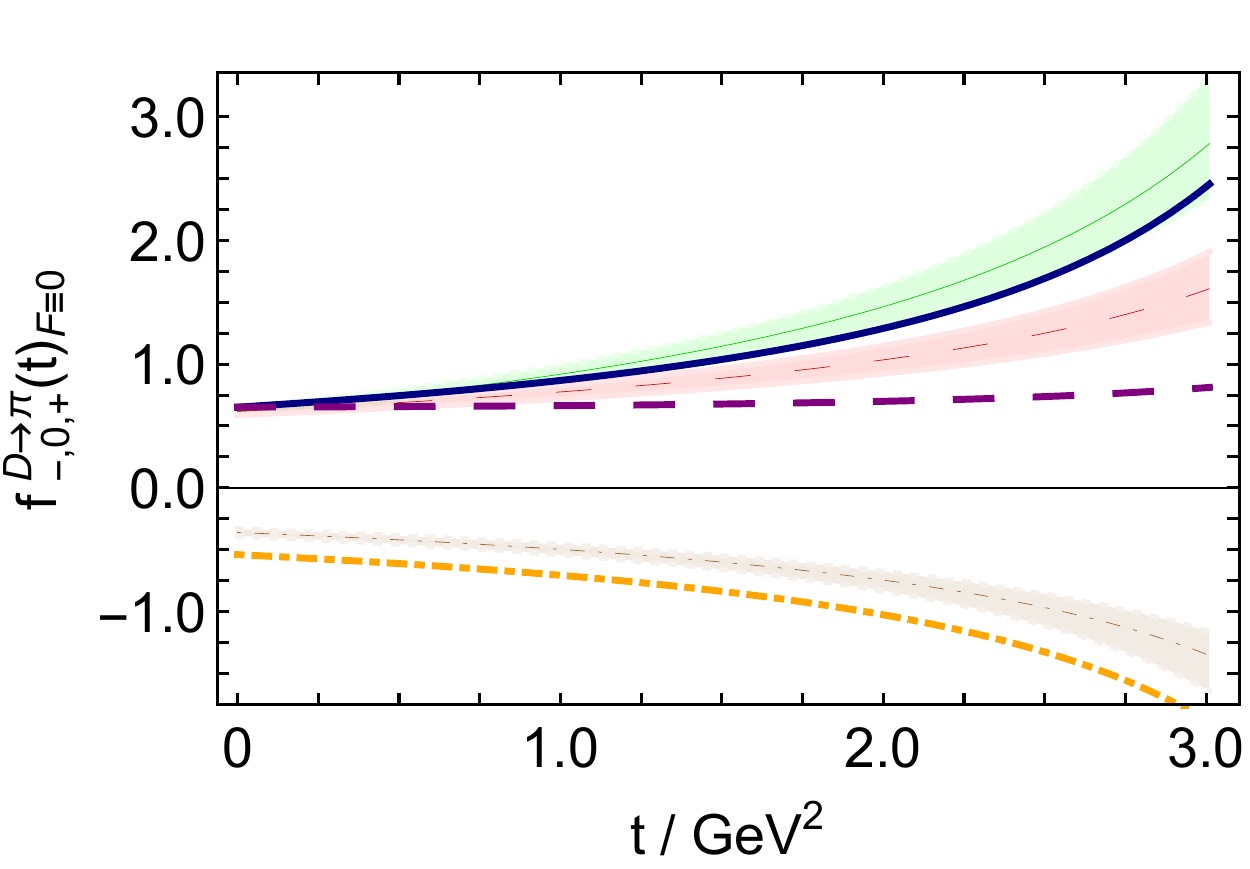}
\caption{\label{FigDpi}
$D \to \pi$ semileptonic transition form factors, defined by Eqs.\,\eqref{dMD}, \eqref{Projections}.
SCI results:
$f_+^{D_u^d}$, solid blue curve; $f_0^{D_u^d}$, long-dashed purple curve; and $f_-^{D_u^d}$, dot-dashed orange curve.
\emph{Upper panel}\,--\,{\sf A} obtained with complete SCI Bethe-Salpeter amplitude in Eq.\,\eqref{PSBSA};
and \emph{lower panel}\,--\,{\sf B}, with $F_{0^-}\equiv 0$ in Eq.\,\eqref{PSBSA} and subsequent repetition of all fitting procedures.
Comparison curves in both panels are drawn from Ref.\,\cite{Yao:2020vef}:
$f_+$, thin solid green curve; $f_0$, long-dashed red curve; and $f_-$, dot-dashed brown curve.  In this and following images, the shaded band around each of the comparison curves indicates the $1\sigma$ confidence level for these predictions.
%, \emph{i.e}.\ 68\% of all SPM approximants lie within the band centred on a given curve.
%
}
\end{figure}

\begin{figure}[t]
\vspace*{2ex}

\leftline{\hspace*{0.5em}{\large{\textsf{A}}}}
\vspace*{-4ex}
\includegraphics[width=0.45\textwidth]{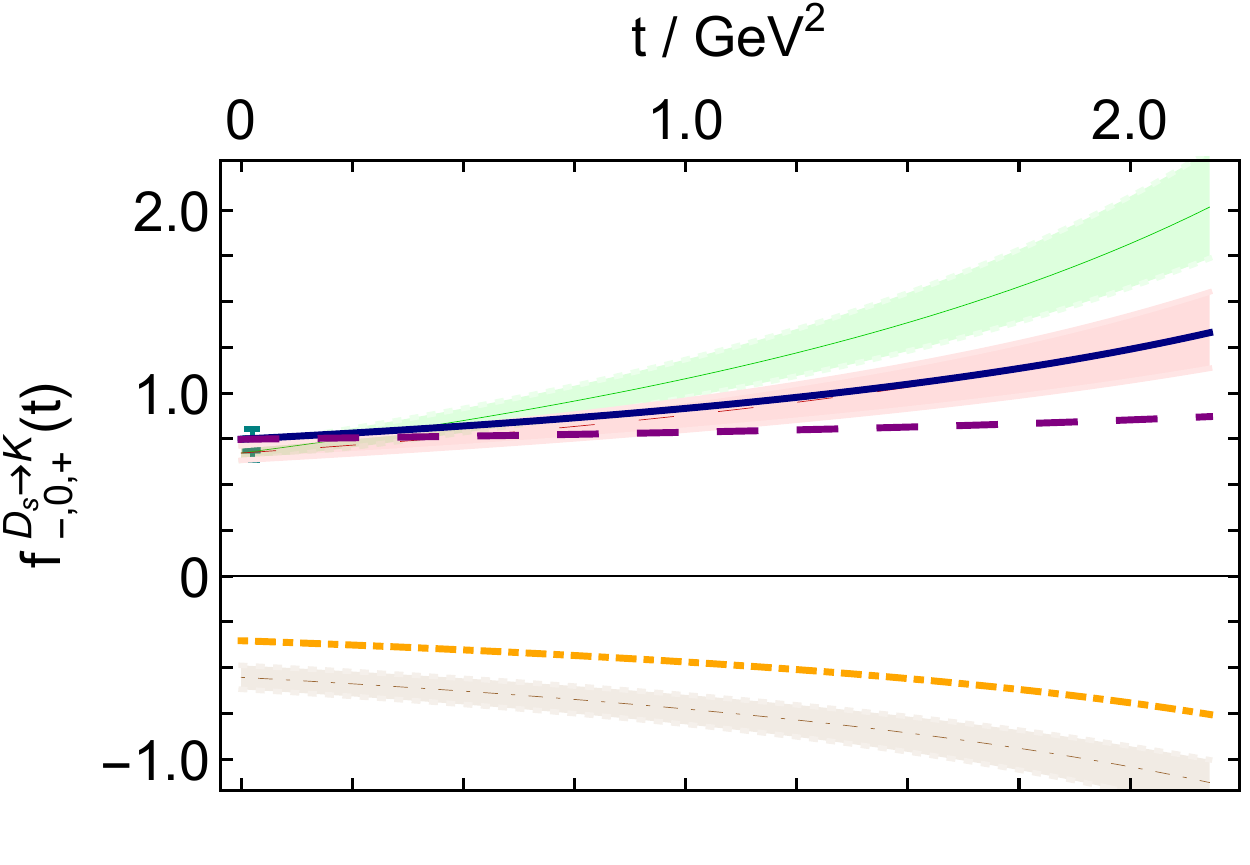}
\vspace*{-2ex}

\leftline{\hspace*{0.5em}{\large{\textsf{B}}}}
\vspace*{-4ex}
\includegraphics[width=0.45\textwidth]{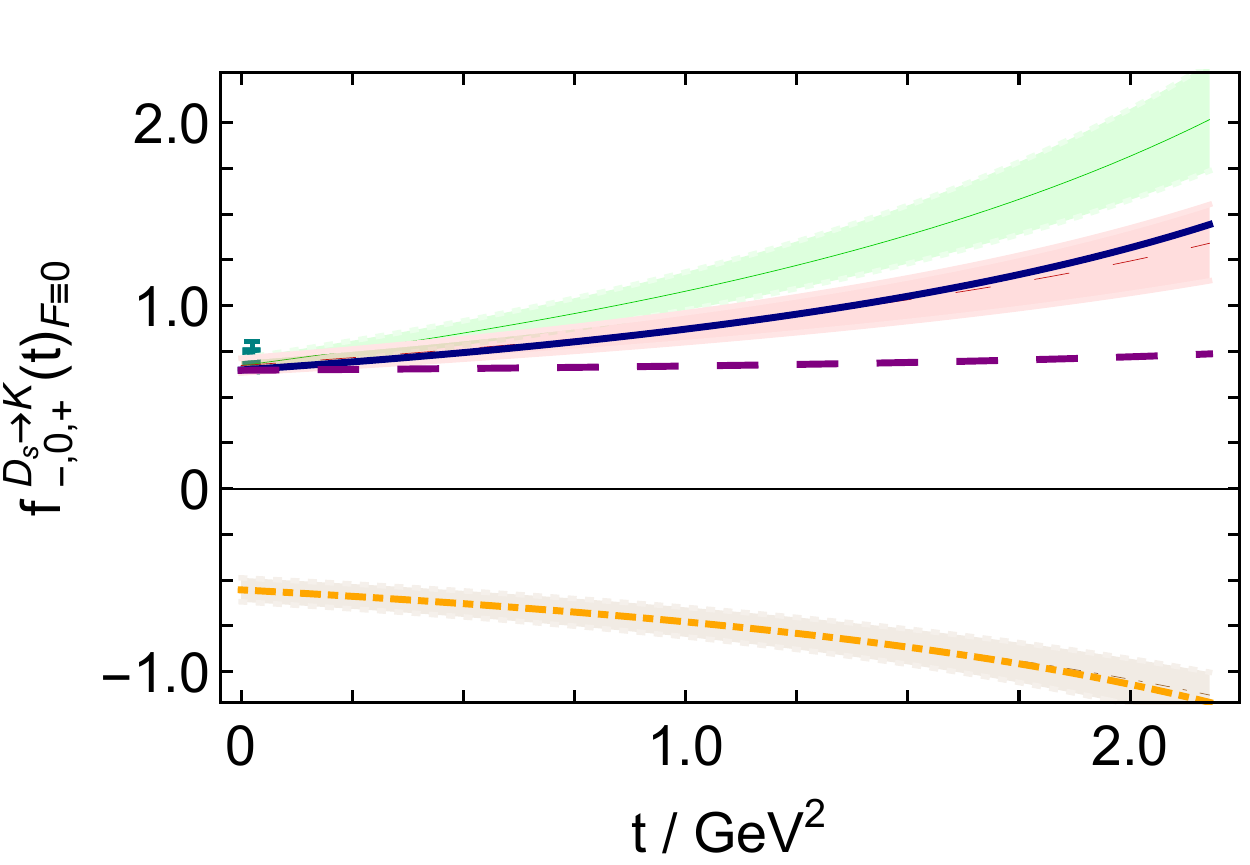}
\caption{\label{FigDsK}
$D_s^+ \to K^0$ transition form factors, defined by analogy with Eqs.\,\eqref{dMD}, \eqref{Projections}.
SCI results:
$f_+^{D_s^d}$, solid blue curve; $f_0^{D_s^d}$, long-dashed purple curve; and $f_-^{D_s^d}$, dot-dashed orange curve.
\emph{Upper panel}\,--\,{\sf A} obtained with complete SCI Bethe-Salpeter amplitude in Eq.\,\eqref{PSBSA};
and \emph{lower panel}\,--\,{\sf B}, with $F_{0^-}\equiv 0$ and subsequent repetition of all fitting procedures.
Comparison curves in both panels are drawn from Ref.\,\cite{Yao:2020vef}:
$f_+$, thin solid green curve; $f_0$, long-dashed red curve; and $f_-$, dot-dashed brown curve.
%The shaded band around each of the comparison curves indicates the $1\sigma$ confidence level for these predictions.
%
Empirical datum, cyan square \cite{Ablikim:2018upe}.
%, \emph{i.e}.\ 68\% of all SPM approximants lie within the band centred on a given curve.
%
}
\end{figure}

\section{Weak $D_{(s)}$ Semileptonic Transitions}
\label{SecSemiLepResults}
%\subsection{Weak $c$-quark Transitions}
%
In the isospin-symmetry limit, there are three distinct such processes: $D^0 \to \pi^-$, $D_s^+ \to K^0$, $D^+ \to \bar K^0$.  The first two measure $c\to d$ and the last, $c\to s$; and all provide information on the environmental sensitivity of these transitions.  These transitions were calculated using a realistic quark+antiquark interaction in Ref.\,\cite{Yao:2020vef}; so, comparison here serves to establish the utility of the SCI as a quantitative guide in such applications.

Results for the maximum-recoil value of the transition form factors are listed in Table~\ref{fp0val}A.  Taken over all quantities listed in the first three rows, the $F_{0^-}\neq 0$ SCI results provide a better match with Ref.\,\cite{Yao:2020vef}.

The SCI $D\to \pi$ transition form factors are depicted in Fig.\,\ref{FigDpi} and compared with analogous form factors drawn from Ref.\,\cite{Yao:2020vef}.  On balance, best agreement is obtained using the SCI results produced with the complete Bethe-Salpeter amplitudes.
It is worth recalling that the $D \to \pi$ semileptonic transition form factors calculated in Ref.\,\cite{Yao:2020vef} are in good agreement with available empirical data \cite{Ablikim:2015ixa}.

A general characteristic of SCI form factors is evident in Fig.\,\ref{FigDpi}; namely, they are typically stiffer than predictions obtained with realistic momentum-depen\-dent interactions \cite{Chen:2012txa}.  This is more of an issue with their spacelike behaviour.  It is ameliorated herein because our focus is the timelike region and, as demonstrated in Sec.\,\ref{WeakVectorVertex}, the SCI preserves the feature that each weak transition vertex possesses a pole at the related meson mass.

The $D_s^+ \to K^0$ transition form factors are drawn in Fig.\,\ref{FigDsK}.  Again, best agreement between the SCI predictions and the results in Ref.\,\cite{Yao:2020vef} is obtained when the complete SCI Bethe-Salpeter amplitude is used.  (We quantified this using a ${\mathpzc L}_1$ measure applied to the independent transition form factors $f_{0,+}$.)  Only one empirical datum is available in this case \cite{Ablikim:2018upe}:
\begin{equation}
 f_+^{D_s^+\to K^0} (0)= 0.720 \pm 0 .084 \, ({\rm stat}) \pm 0.013 \, ({\rm syst})\,.
\end{equation}
It is plotted in Fig.\,\ref{FigDsK} and agrees with the SCI result obtained using the complete Bethe-Salpeter amplitude, Fig.\,\ref{FigDsK}A.

\begin{figure}[t]
\vspace*{2ex}

\leftline{\hspace*{0.5em}{\large{\textsf{A}}}}
\vspace*{-4ex}
\includegraphics[width=0.45\textwidth]{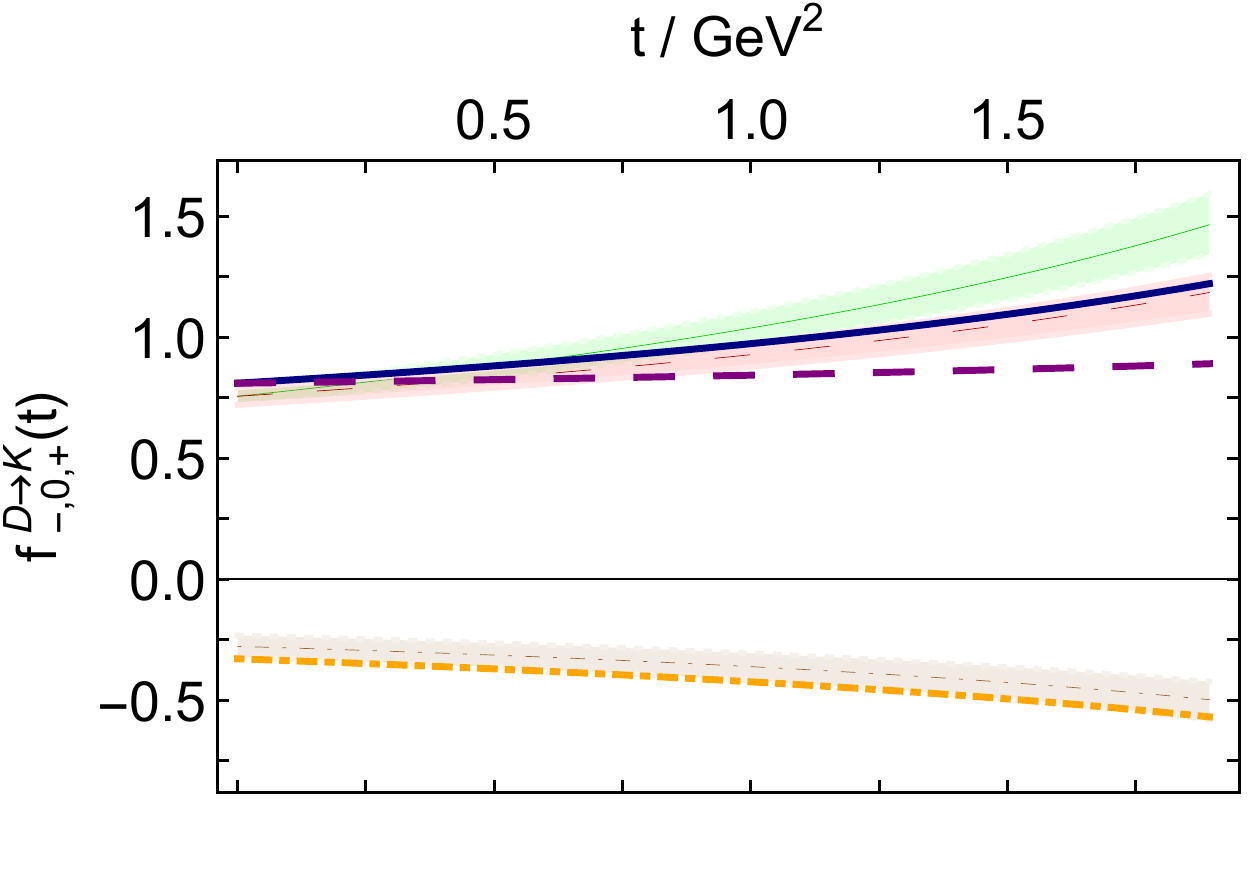}
\vspace*{-2ex}

\leftline{\hspace*{0.5em}{\large{\textsf{B}}}}
\vspace*{-4ex}
\includegraphics[width=0.45\textwidth]{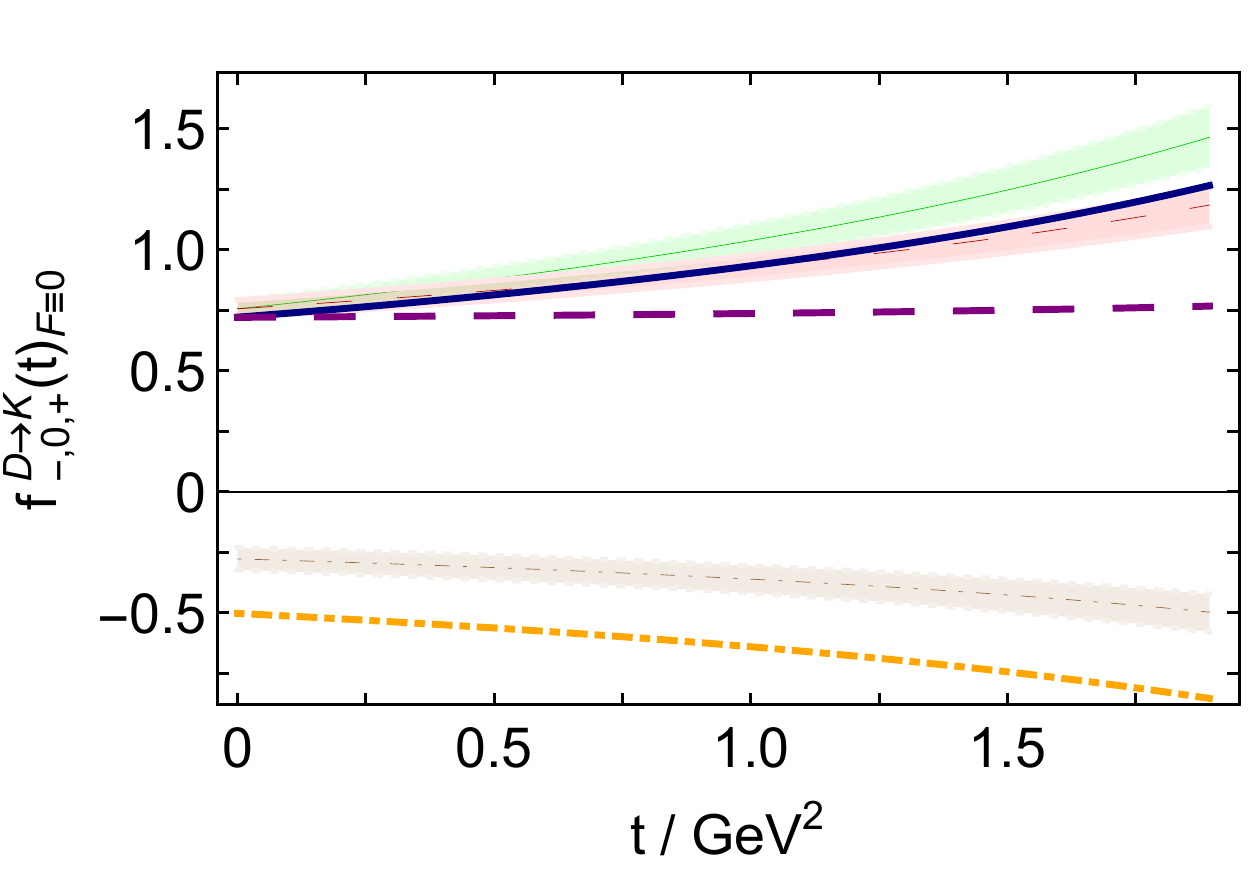}
\caption{\label{FigDK}
$D \to K$ transition form factors, defined by analogy with Eqs.\,\eqref{dMD}, \eqref{Projections}.
SCI results:
$f_+^{D_d^s}$, solid blue curve; $f_0^{D_d^s}$, long-dashed purple curve; and $f_-^{D_d^s}$, dot-dashed orange curve.
\emph{Upper panel}\,--\,{\sf A} obtained with complete SCI Bethe-Salpeter amplitude in Eq.\,\eqref{PSBSA};
and \emph{lower panel}\,--\,{\sf B}, with $F_{0^-}\equiv 0$ and subsequent repetition of all fitting procedures.
Comparison curves in both panels are drawn from Ref.\,\cite{Yao:2020vef}:
$f_+$, thin solid green curve; $f_0$, long-dashed red curve; and $f_-$, dot-dashed brown curve.  %The shaded band around each of the comparison curves indicates the $1\sigma$ confidence level for these predictions.
}
\end{figure}

The $D\to K$ transition form factors are drawn in Fig.\,\ref{FigDK}.  Best agreement with the predictions from Ref.\,\cite{Yao:2020vef} is obtained using the complete SCI Bethe-Salpeter amplitude, \emph{i.e}.\ with the $F_{0^-} \neq 0$ values listed in Table~\ref{Tab:MesonSpectrum}.
The $D \to K$ semileptonic transition form factors in Ref.\,\cite{Yao:2020vef} agree well with available empirical data \cite{Ablikim:2015ixa}.

%\subsection{Branching Fractions}
%\label{SecBF}
%
With computed transition form factors in hand, one can calculate partial widths for the associated decays.  Remaining with $D\to \pi$ as our exemplar, the relevant expression is \cite{Bajc:1995km}:
\begin{subequations}
\label{DecayFraction}
\begin{align}
\Gamma_{D^0 \pi^-} & = |V_{cd}|^2 \frac{G_F^2  m_{D}^2}{24 \pi^3}\nonumber \\
& \quad \times
\int_0^{y_m^{D \pi}} \!\! dy\,[f_+^{D_u^d}(y \, m_{D}^2) ]^2 k_{D \pi}^3(y)\,,\\
k_{D \pi}^2(t) & = (m_{D}^2 (1-y) + m_\pi^2)^2/[4 m_{D}^2] - m_\pi^2,
\end{align}
\end{subequations}
with $G_F = 1.166 \times 10^{-5}\,$GeV$^{-2}$.  Using the SCI results for $f_+^{D_u^d}(t)$, the associated branching fractions are listed in Table~\ref{fp0val}B: the $F_{0^-}$ result for this integrated quantity agrees well with other theory \cite{Yao:2020vef} and experiment \cite{Ablikim:2015ixa}.

Analogous SCI results for $D_s^+ \to K^0 e^+ \nu_e$, $D^0 \to K^- e^+ \nu_e$ branching fractions are also listed in Table~\ref{fp0val}B.  Again, those obtained with the complete Bethe-Salpeter amplitude compare favourably with other theory \cite{Yao:2020vef} and experiment \cite{Ablikim:2015ixa, Ablikim:2018upe}.

\section{Weak $B_{(c,s)}$ Semileptonic Transitions}
\label{BWSLT}
%\subsection{Weak $b$-quark Transitions}
%
Here in the isospin-symmetry limit, there are seven distinct processes, which we choose to be:
%
% ... Vub
(\emph{i}) $\bar B^0 \to \pi^+$,
$B_s \to K$;
% ... Vcb
(\emph{ii}) $B^0 \to D^-$,
$B_s^0 \to D_s^-$,
$B_c^- \to \eta_c$;
% ... Vcd
(\emph{iii}) $B_c^+ \to B^0$;
% ... Vcs
and (\emph{iv}) $B_c^+ \to B_s^0$.
The first two measure $b\to u$ (\emph{i});
the next three, $b\to c$ (\emph{ii});
the sixth, $c\to d$ (\emph{iii}); and the last, $c\to s$ (\emph{iv}); plus, naturally, their environmental sensitivity.  In these cases, no continuum study with the character of Ref.\,\cite{Yao:2020vef} is available; but lQCD methods have been used by many groups, \emph{e.g}.\ Refs.\,\cite{McLean:2019qcx, Colquhoun:2016osw, Cooper:2020wnj, Bailey:2008wp, AlHaydari:2009zr}, and numerous model studies exist, \emph{e.g}.\ Refs.\,\cite{Melikhov:1997wp, Melikhov:2001zv, Faessler:2002ut, Ebert:2003wc, Ball:2004ye, Wu:2006rd, Khodjamirian:2006st, Zhao:2006at, Lu:2007sg, Ivanov:2007cw, Faustov:2014bxa, Hu:2019qcn, Hu:2019bdf, Choi:2021mni}.

\begin{figure}[t]
\vspace*{2ex}

\leftline{\hspace*{0.5em}{\large{\textsf{A}}}}
\vspace*{-4ex}
\includegraphics[width=0.45\textwidth]{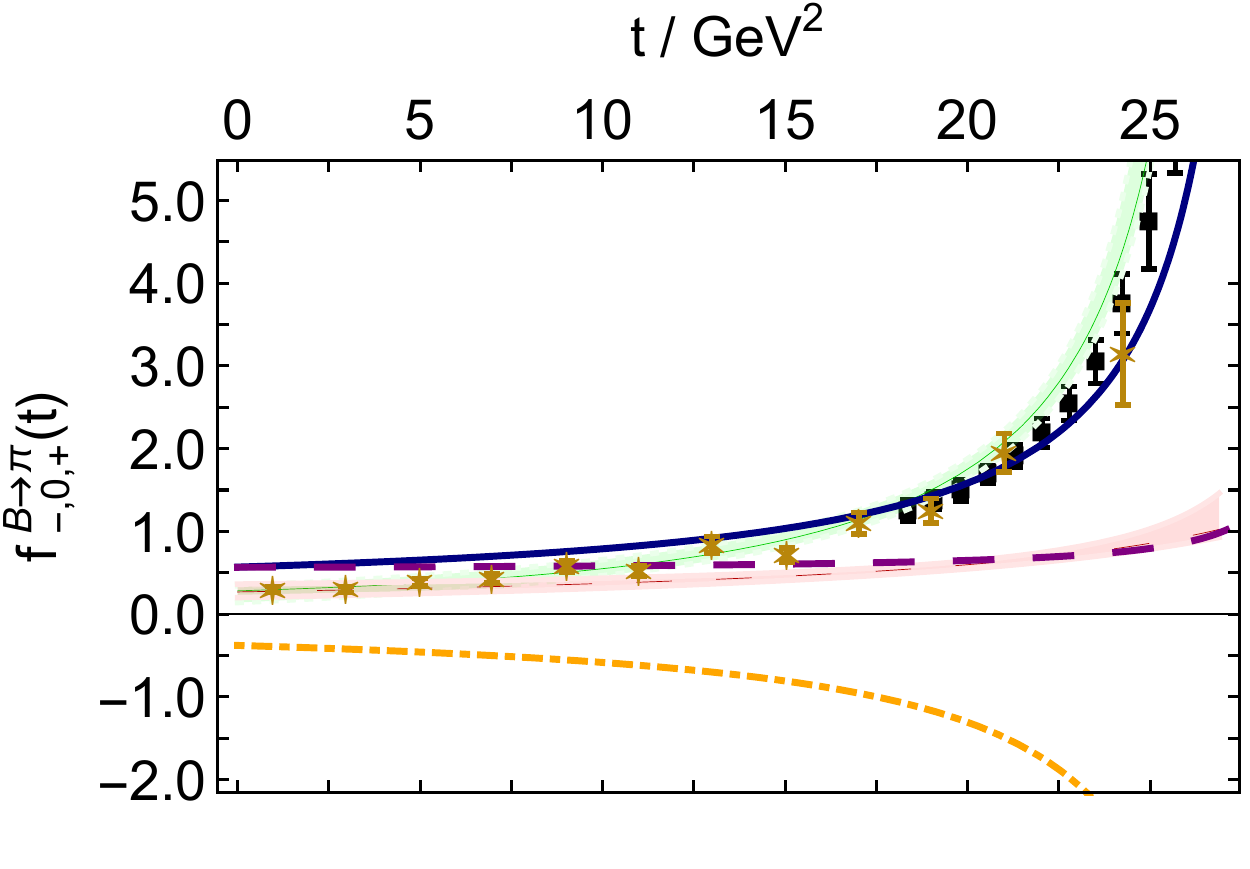}
\vspace*{-2ex}

\leftline{\hspace*{0.5em}{\large{\textsf{B}}}}
\vspace*{-4ex}
\includegraphics[width=0.45\textwidth]{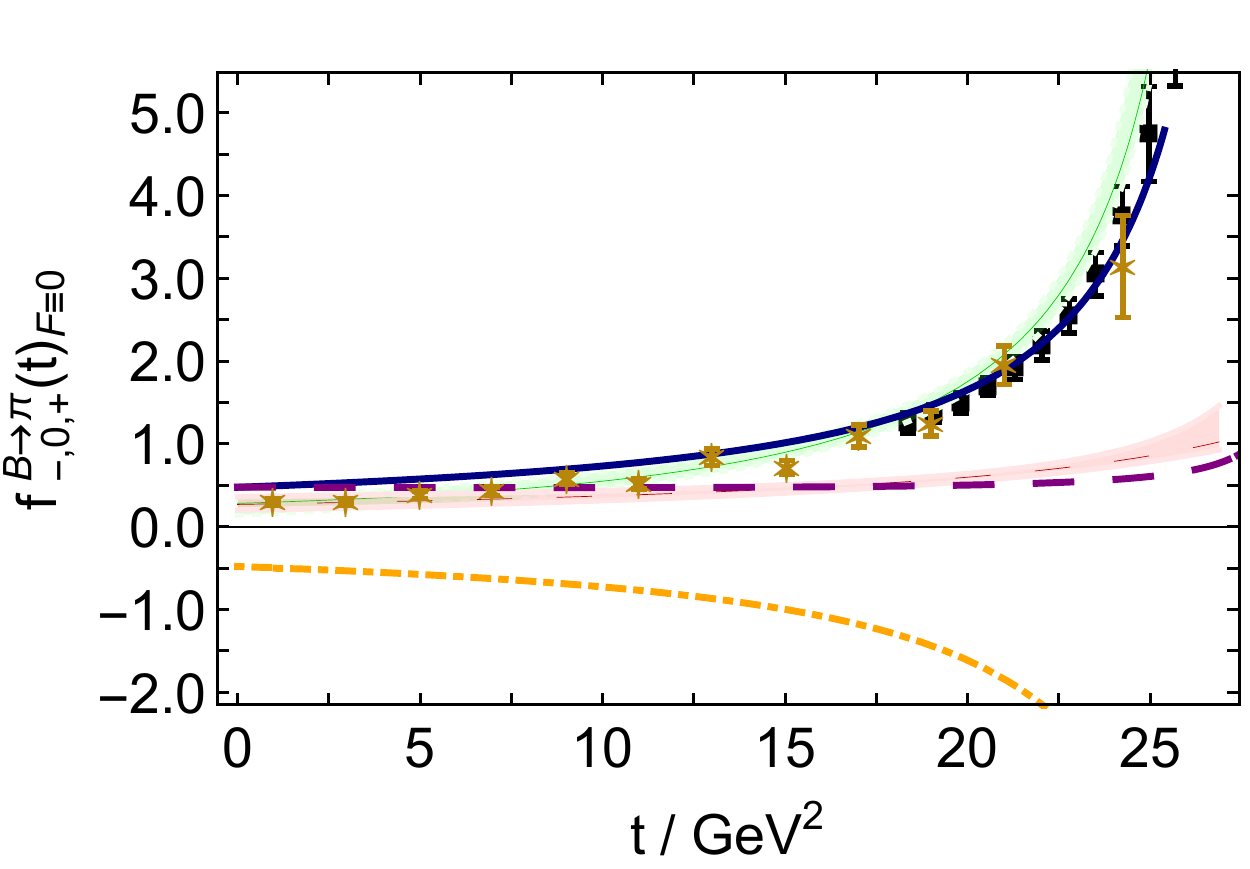}
\caption{\label{FigBpi}
$B \to \pi$ semileptonic transition form factors, defined by analogy with Eqs.\,\eqref{dMD}, \eqref{Projections}.
SCI results:
$f_+^{\bar B^u_{d}}$, solid blue curve; $f_0^{\bar B^u_{d}}$, long-dashed purple curve; and $f_-^{\bar B^u_{d}}$, dot-dashed orange curve.
\emph{Upper panel}\,--\,{\sf A} obtained with complete SCI Bethe-Salpeter amplitude in Eq.\,\eqref{PSBSA};
and \emph{lower panel}\,--\,{\sf B}, with $F_{0^-}\equiv 0$ and subsequent refitting.
In both panels, the data (gold stars) are from the BaBar Collaboration \cite{Aubert:2006px};
the other points (black squares) are lQCD results obtained with $2+1$ flavours of light sea quarks \cite{Bailey:2008wp};
and the comparison bands are drawn from Ref.\,\cite{AlHaydari:2009zr}, being fits to quenched-lQCD calculations.
}
\end{figure}

\begin{figure}[t]
\vspace*{2ex}

\leftline{\hspace*{0.5em}{\large{\textsf{A}}}}
\vspace*{-4ex}
\includegraphics[width=0.45\textwidth]{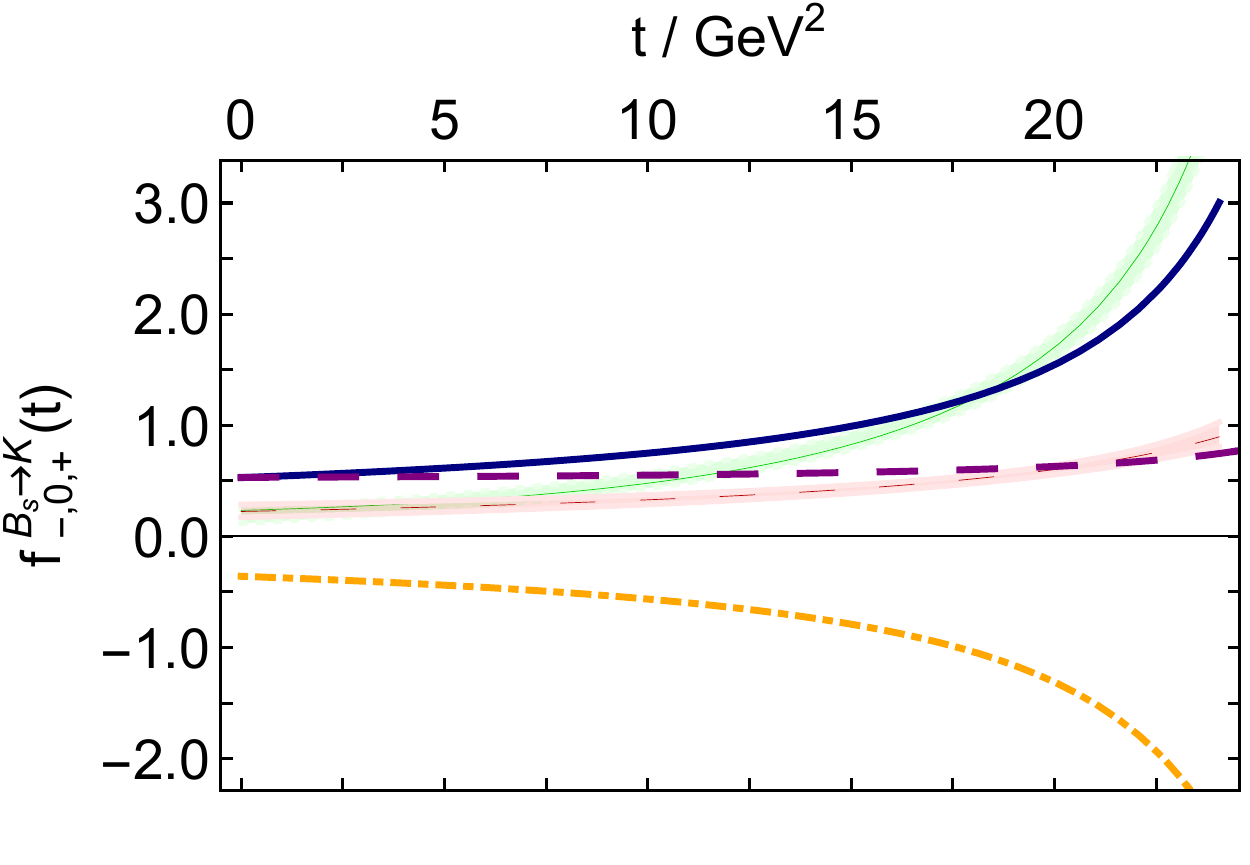}
\vspace*{-2ex}

\leftline{\hspace*{0.5em}{\large{\textsf{B}}}}
\vspace*{-4ex}
\includegraphics[width=0.45\textwidth]{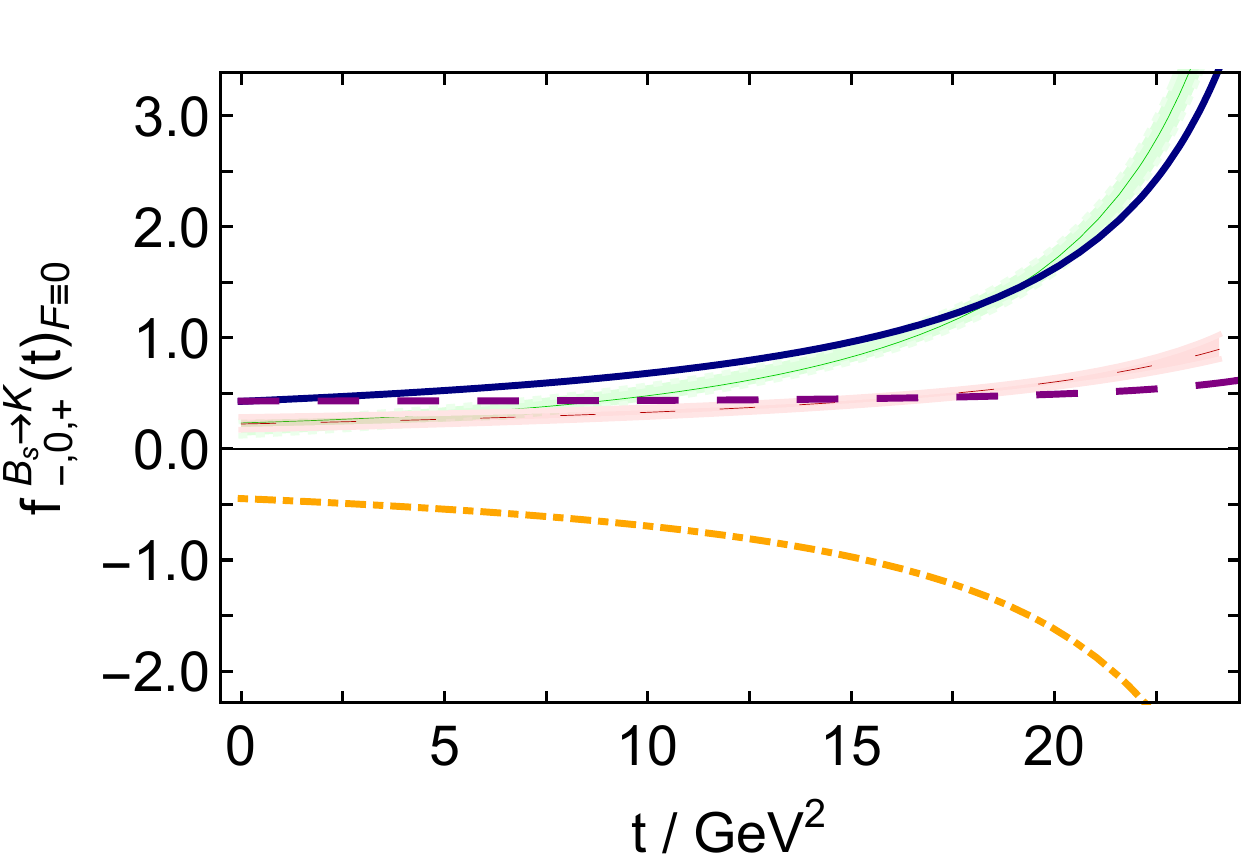}
\caption{\label{FigBsK}
$B_s \to K$ transition form factors, defined by analogy with Eqs.\,\eqref{dMD}, \eqref{Projections}.
SCI results:
$f_+^{B_{s}^{u}}$, solid blue curve; $f_0^{B_{s}^{u}}$, long-dashed purple curve; and $f_-^{B_{s}^{u}}$, dot-dashed orange curve.
\emph{Upper panel}\,--\,{\sf A} obtained with complete SCI Bethe-Salpeter amplitude in Eq.\,\eqref{PSBSA};
and \emph{lower panel}\,--\,{\sf B}, with $F_{0^-}\equiv 0$ and subsequent repetition of all fitting procedures.
In both panels, the comparison bands are drawn from Ref.\,\cite{AlHaydari:2009zr}, being fits to quenched-lQCD calculations.
}
\end{figure}

\begin{figure}[t]
\vspace*{2ex}

\leftline{\hspace*{0.5em}{\large{\textsf{A}}}}
\vspace*{-4ex}
\includegraphics[width=0.45\textwidth]{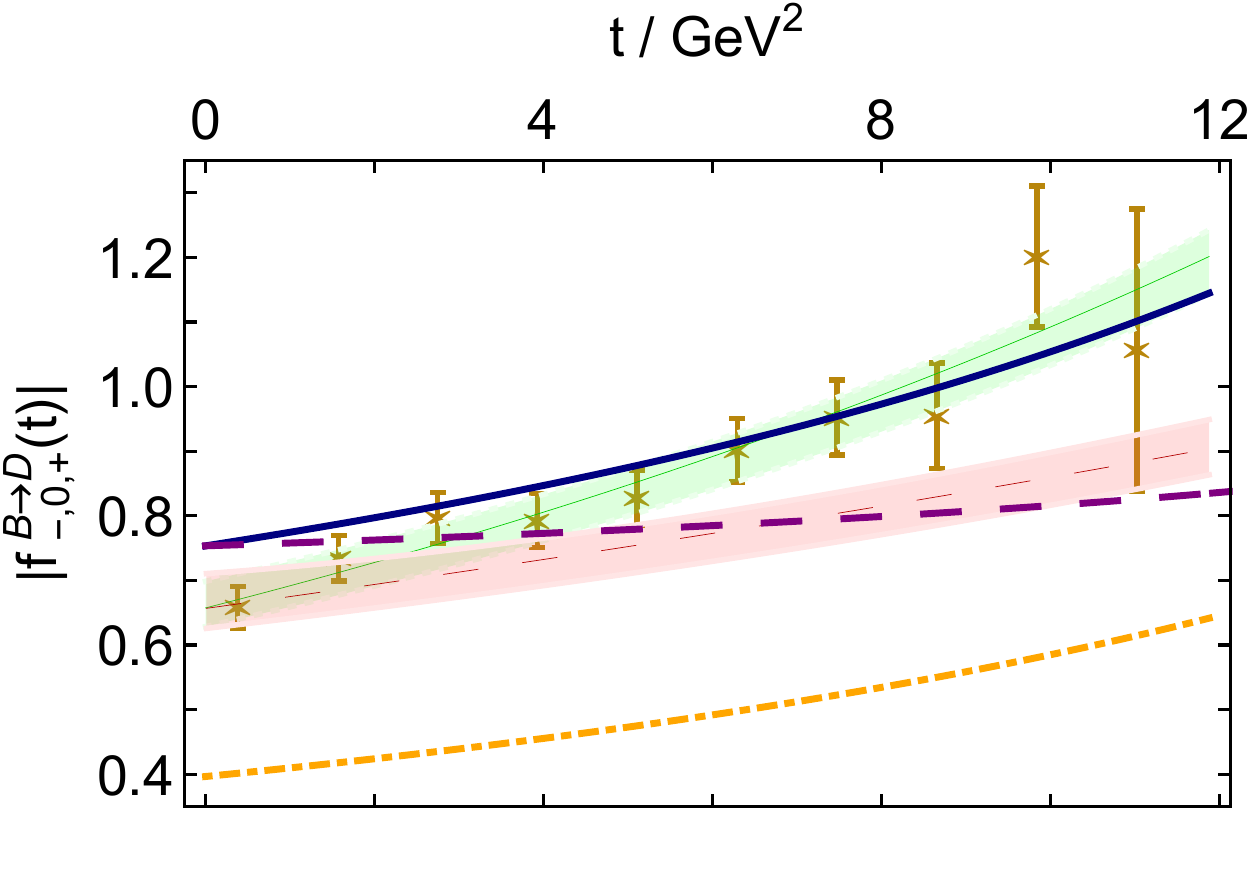}
\vspace*{-2ex}

\leftline{\hspace*{0.5em}{\large{\textsf{B}}}}
\vspace*{-4ex}
\includegraphics[width=0.45\textwidth]{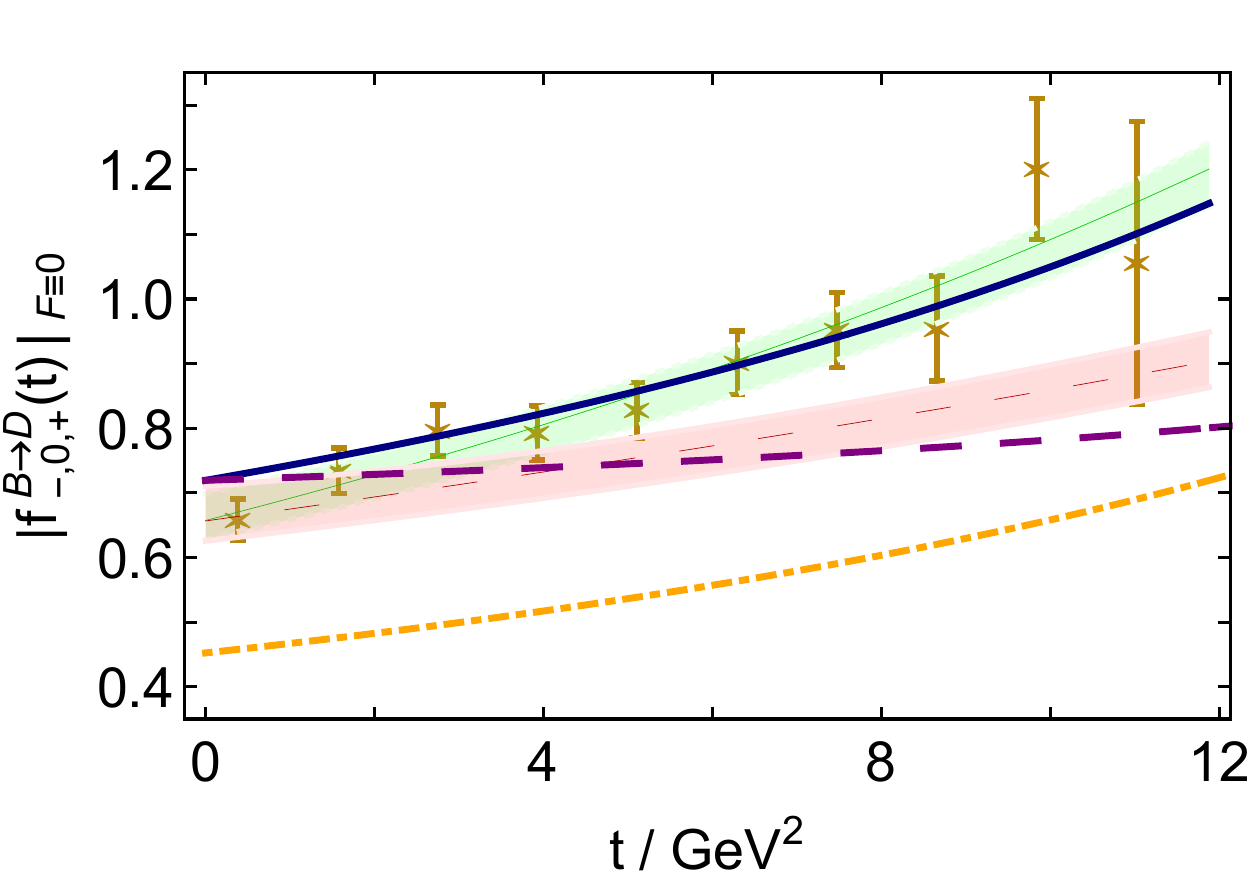}
\caption{\label{FigBD}
$B\to D $ transition form factors, defined by analogy with Eqs.\,\eqref{dMD}, \eqref{Projections}.
SCI results:
$f_+^{B_{u}^{\bar c}}$, solid blue curve; $f_0^{B_{u}^{\bar c}}$, long-dashed purple curve; and $-f_-^{B_{u}^{\bar c}}$, dot-dashed orange curve.
\emph{Upper panel}\,--\,{\sf A} obtained with complete SCI Bethe-Salpeter amplitude in Eq.\,\eqref{PSBSA};
and \emph{lower panel}\,--\,{\sf B}, with $F_{0^-}\equiv 0$ and subsequent repetition of all fitting procedures.
In both panels, the data (gold stars, $f_+$) are from the BaBar Collaboration \cite{Aubert:2009ac}; and
the comparison bands (thin dashed red, $f_0$; and thin green, $f_+$) are drawn from Ref.\,\cite{Yao:2019vty}, being parametrisations built from recent experimental data and lQCD results.
}
\end{figure}

\begin{figure}[t]
\vspace*{2ex}

\leftline{\hspace*{0.5em}{\large{\textsf{A}}}}
\vspace*{-4ex}
\includegraphics[width=0.45\textwidth]{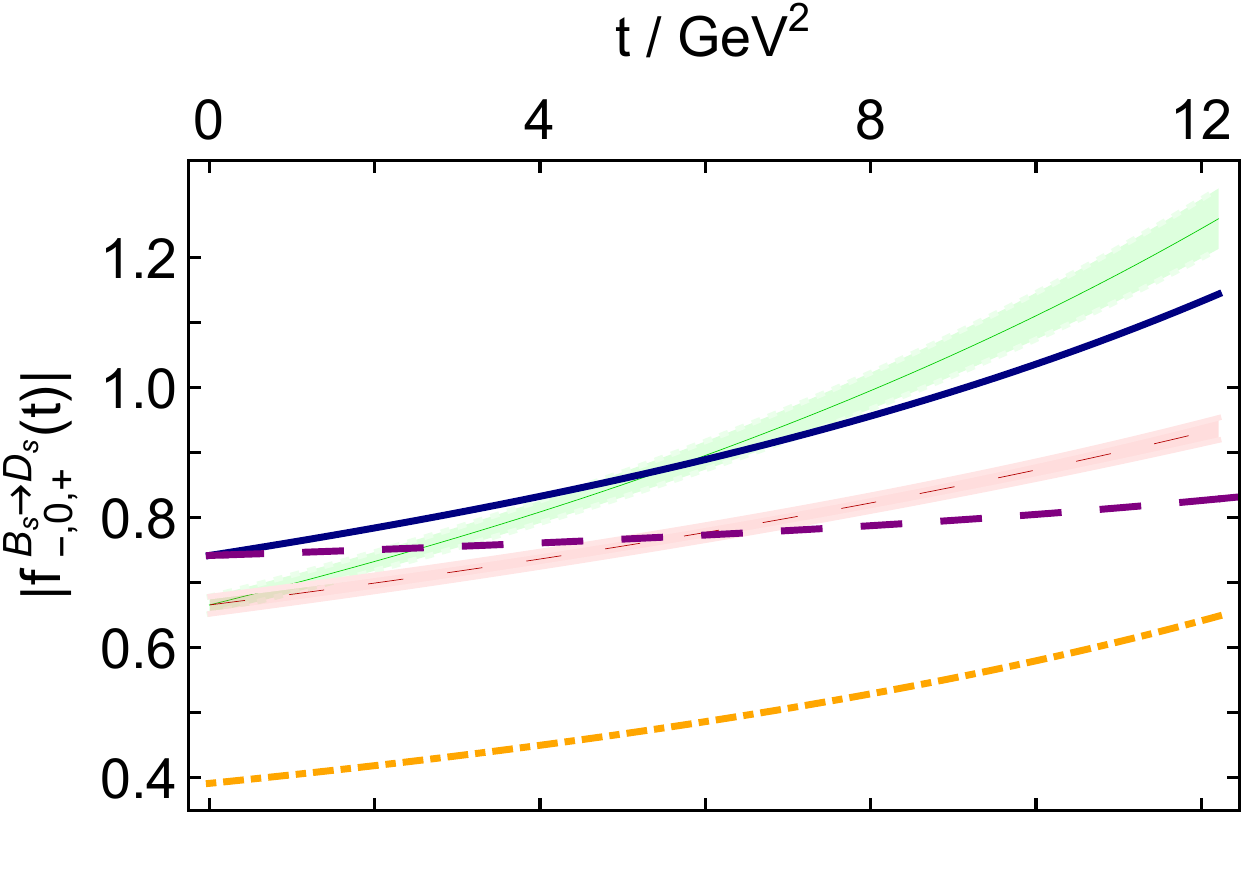}
\vspace*{-2ex}

\leftline{\hspace*{0.5em}{\large{\textsf{B}}}}
\vspace*{-4ex}
\includegraphics[width=0.45\textwidth]{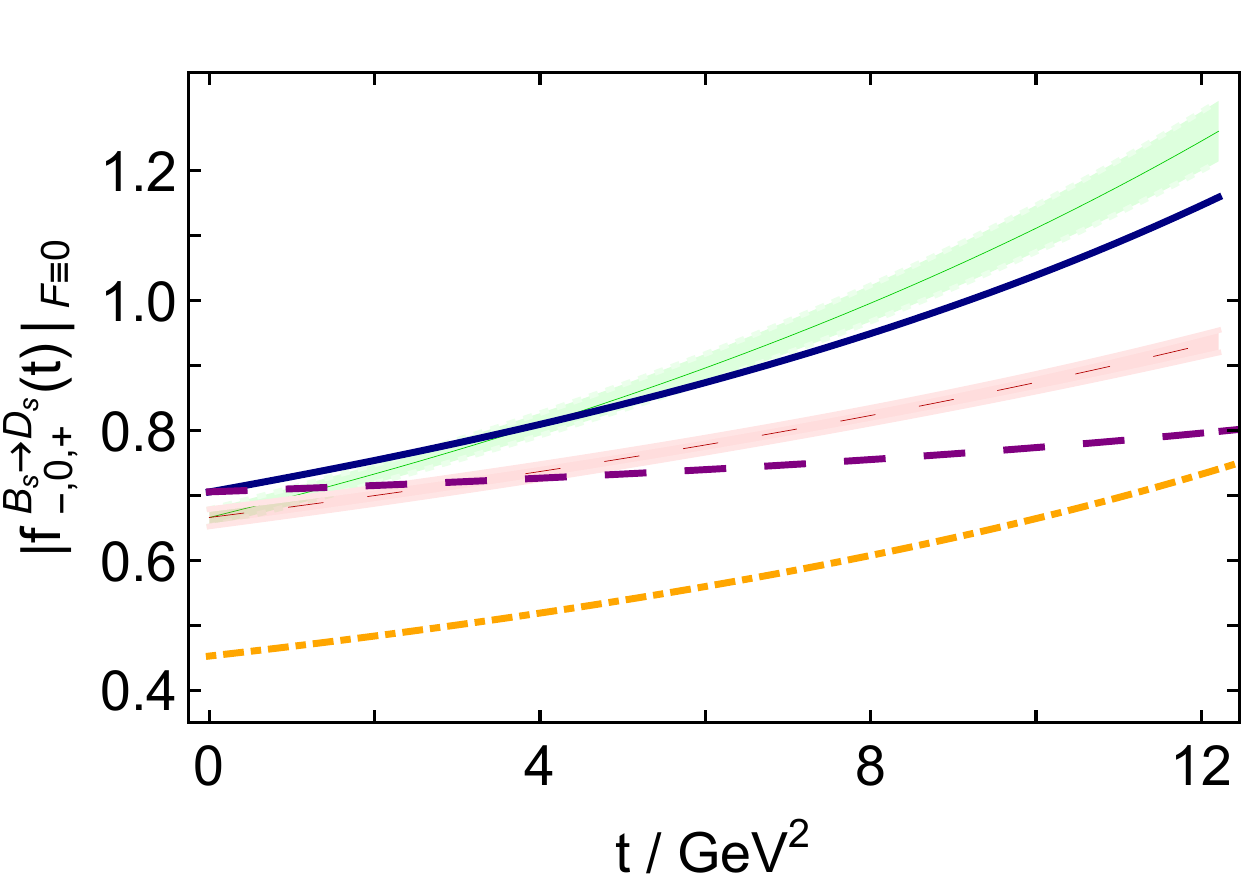}
\caption{\label{FigBsDs}
$B_s\to D_s $ transition form factors, defined by analogy with Eqs.\,\eqref{dMD}, \eqref{Projections}.
SCI results:
$f_+^{B_{s}^{\bar c}}$, solid blue curve; $f_0^{B_{s}^{\bar c}}$, long-dashed purple curve; and $-f_-^{B_{s}^{\bar c}}$, dot-dashed orange curve.
\emph{Upper panel}\,--\,{\sf A} obtained with complete SCI Bethe-Salpeter amplitude in Eq.\,\eqref{PSBSA};
and \emph{lower panel}\,--\,{\sf B}, with $F_{0^-}\equiv 0$ and subsequent refitting.
In both panels, the comparison bands (thin dashed red, $f_0$; and thin green, $f_+$) are drawn from Ref.\,\cite{McLean:2019qcx}, being lQCD results obtained with inclusion of $2+1+1$ flavours of sea quark.
}
\end{figure}

\begin{figure}[t]
\vspace*{2ex}

\leftline{\hspace*{0.5em}{\large{\textsf{A}}}}
\vspace*{-4ex}
\includegraphics[width=0.45\textwidth]{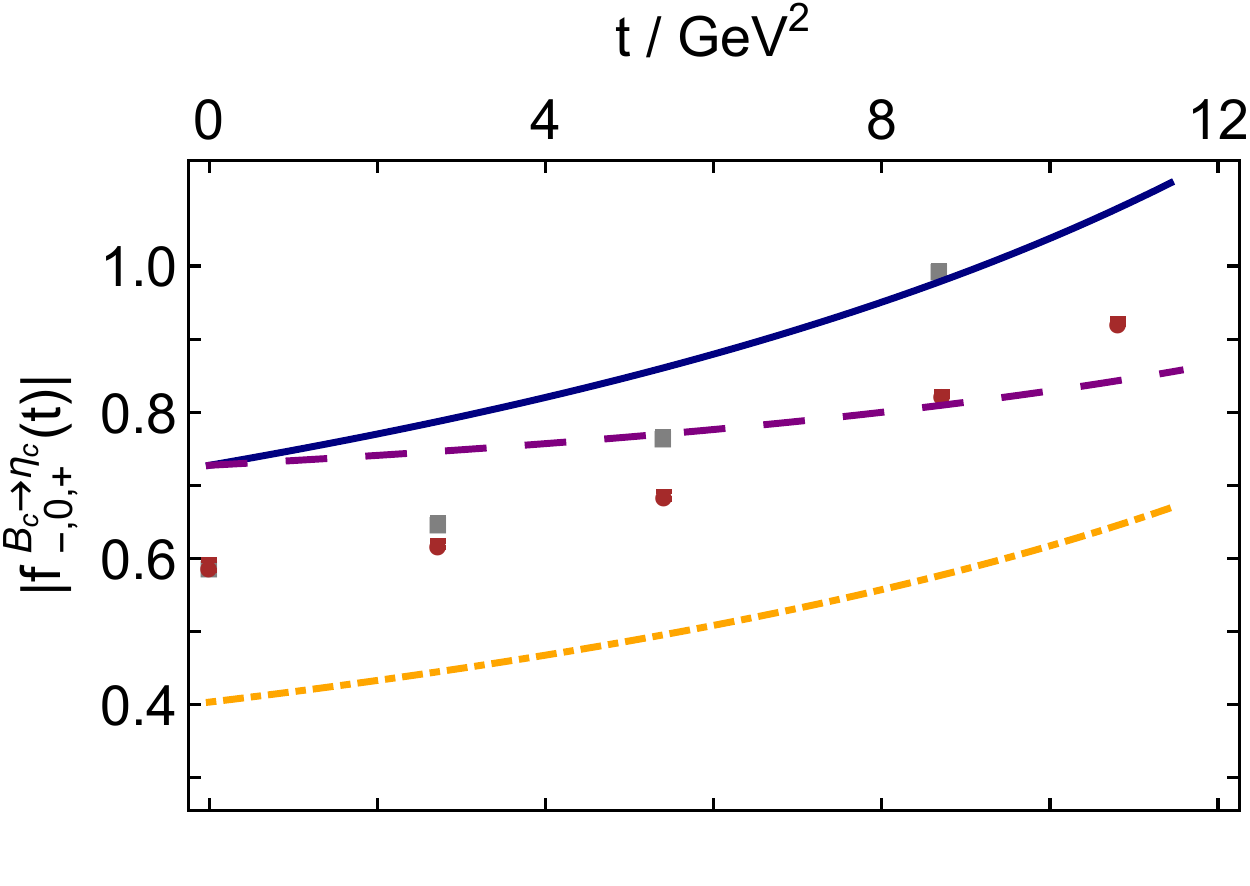}
\vspace*{-2ex}

\leftline{\hspace*{0.5em}{\large{\textsf{B}}}}
\vspace*{-4ex}
\includegraphics[width=0.45\textwidth]{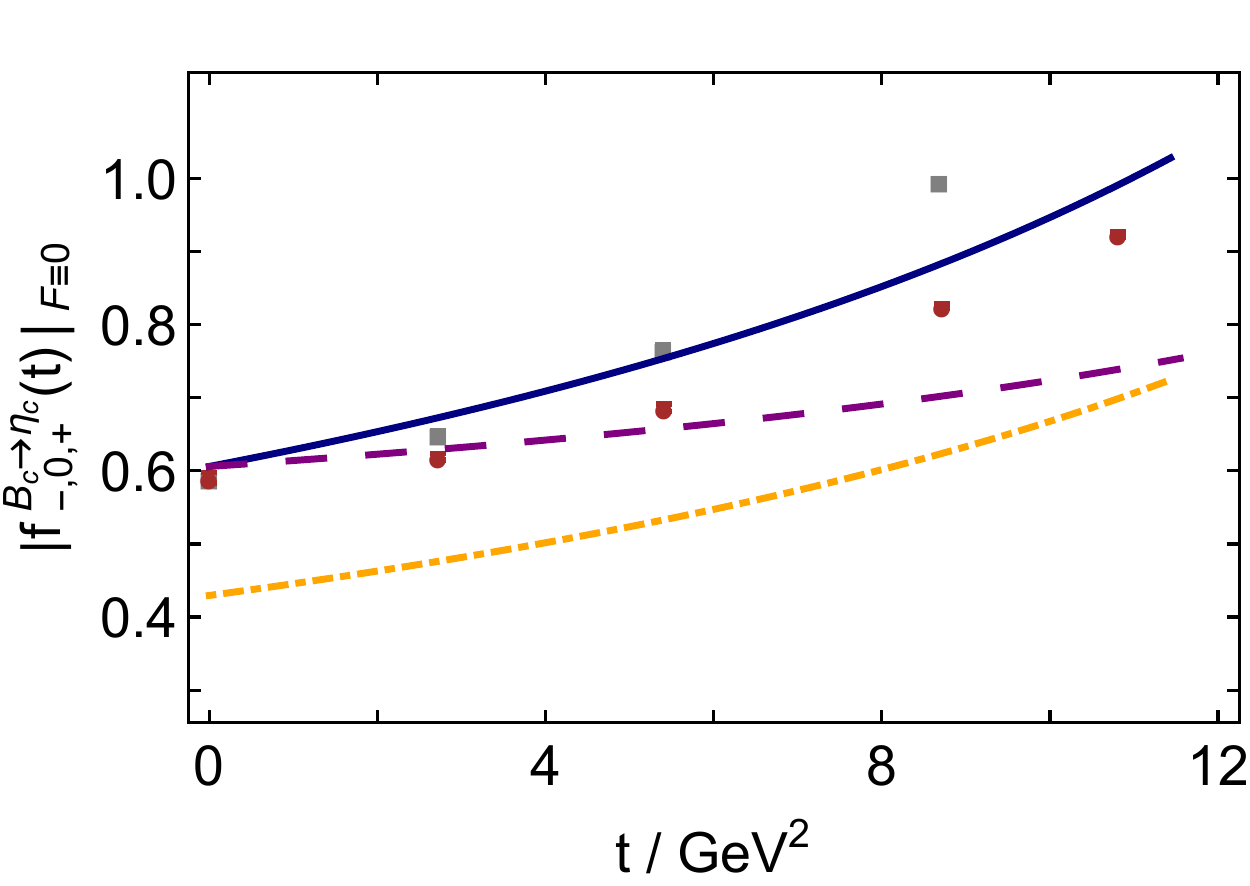}
\caption{\label{FigBetac}
$B_c\to \eta_c $ transition form factors, defined by analogy with Eqs.\,\eqref{dMD}, \eqref{Projections}.
SCI results:
$f_+^{B_{c}^{\bar c}}$, solid blue curve; $f_0^{B_{c}^{\bar c}}$, long-dashed purple curve; and $-f_-^{B_{c}^{\bar c}}$, dot-dashed orange curve.
\emph{Upper panel}\,--\,{\sf A} obtained with complete SCI Bethe-Salpeter amplitude in Eq.\,\eqref{PSBSA};
and \emph{lower panel}\,--\,{\sf B}, with $F_{0^-}\equiv 0$ and subsequent refitting.
In both panels, the comparison points (brown circles, $f_0$; and grey squares, $f_+$) are drawn from Ref.\,\cite{Colquhoun:2016osw}, being exploratory lQCD results.
}
\end{figure}

\begin{figure}[t]
\vspace*{2ex}

\leftline{\hspace*{0.5em}{\large{\textsf{A}}}}
\vspace*{-4ex}
\includegraphics[width=0.45\textwidth]{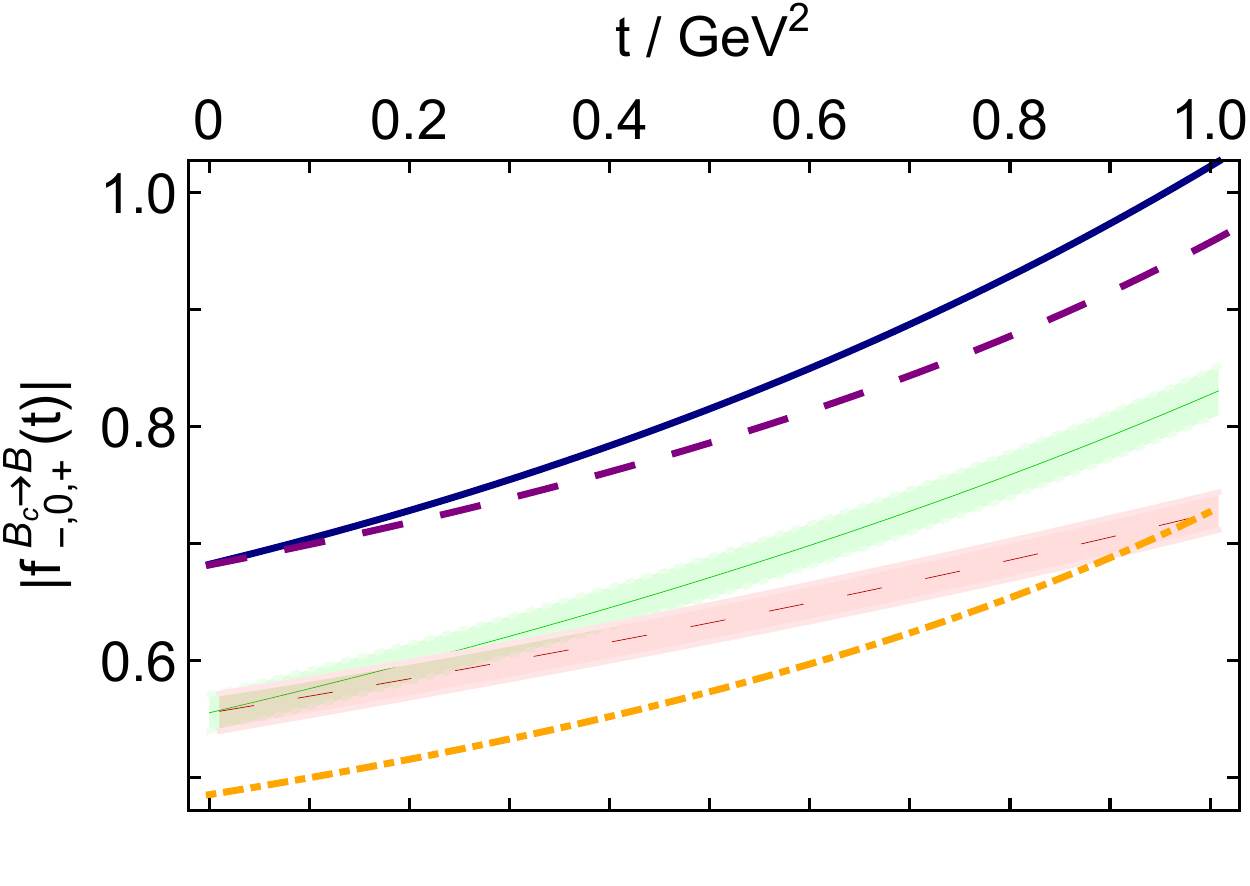}
\vspace*{-2ex}

\leftline{\hspace*{0.5em}{\large{\textsf{B}}}}
\vspace*{-4ex}
\includegraphics[width=0.45\textwidth]{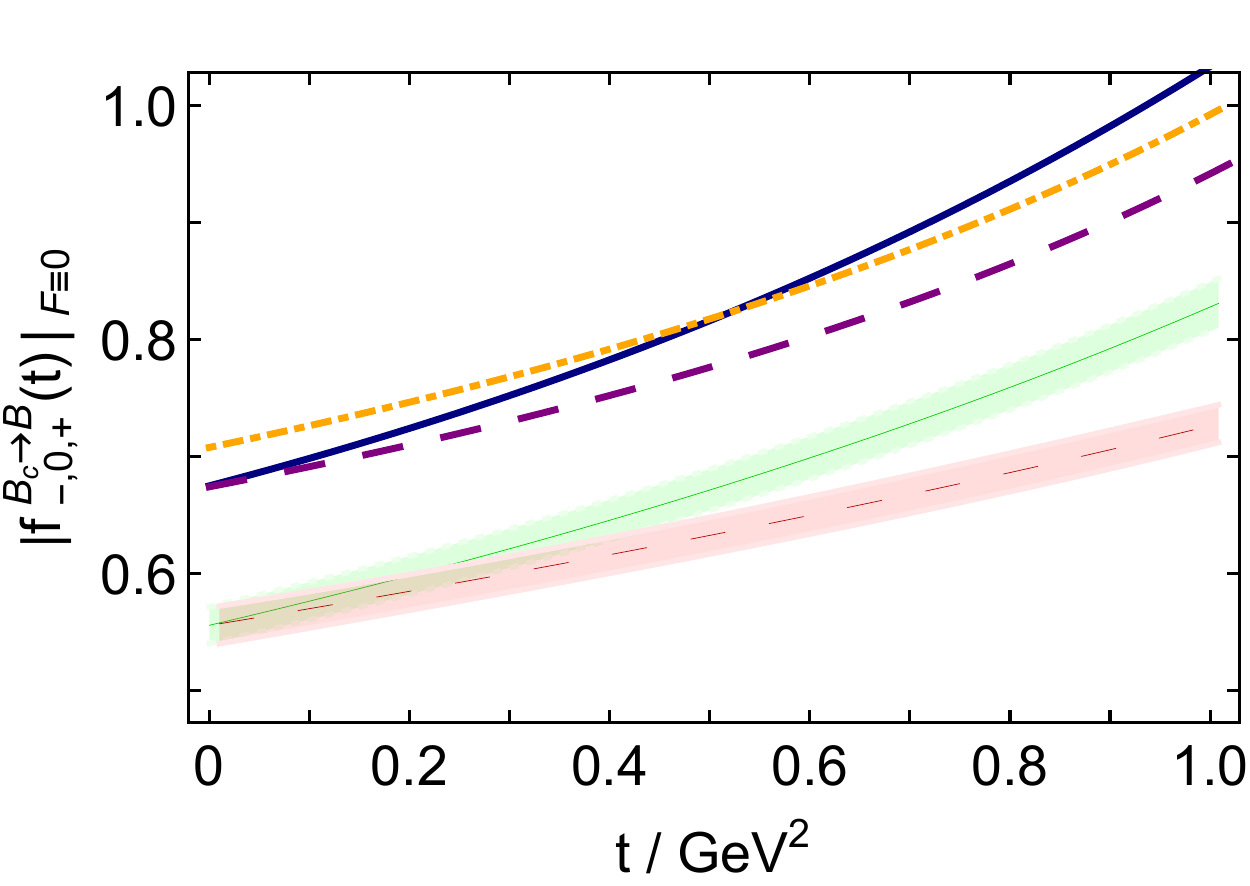}
\caption{\label{FigBcB}
$B_c\to B $ transition form factors, defined by analogy with Eqs.\,\eqref{dMD}, \eqref{Projections}.
SCI results:
$f_+^{B_{b}^{d}}$, solid blue curve; $f_0^{B_{b}^{d}}$, long-dashed purple curve; and $-f_-^{B_{b}^{d}}$, dot-dashed orange curve.
\emph{Upper panel}\,--\,{\sf A} obtained with complete SCI Bethe-Salpeter amplitude in Eq.\,\eqref{PSBSA};
and \emph{lower panel}\,--\,{\sf B}, with $F_{0^-}\equiv 0$  and subsequent refitting.
In both panels, the comparison bands (thin dashed red, $f_0$; and thin green, $f_+$) are lQCD results from  Ref.\,\cite{Cooper:2020wnj}.
}
\end{figure}

\begin{figure}[t]
\vspace*{2ex}

\leftline{\hspace*{0.5em}{\large{\textsf{A}}}}
\vspace*{-4ex}
\includegraphics[width=0.45\textwidth]{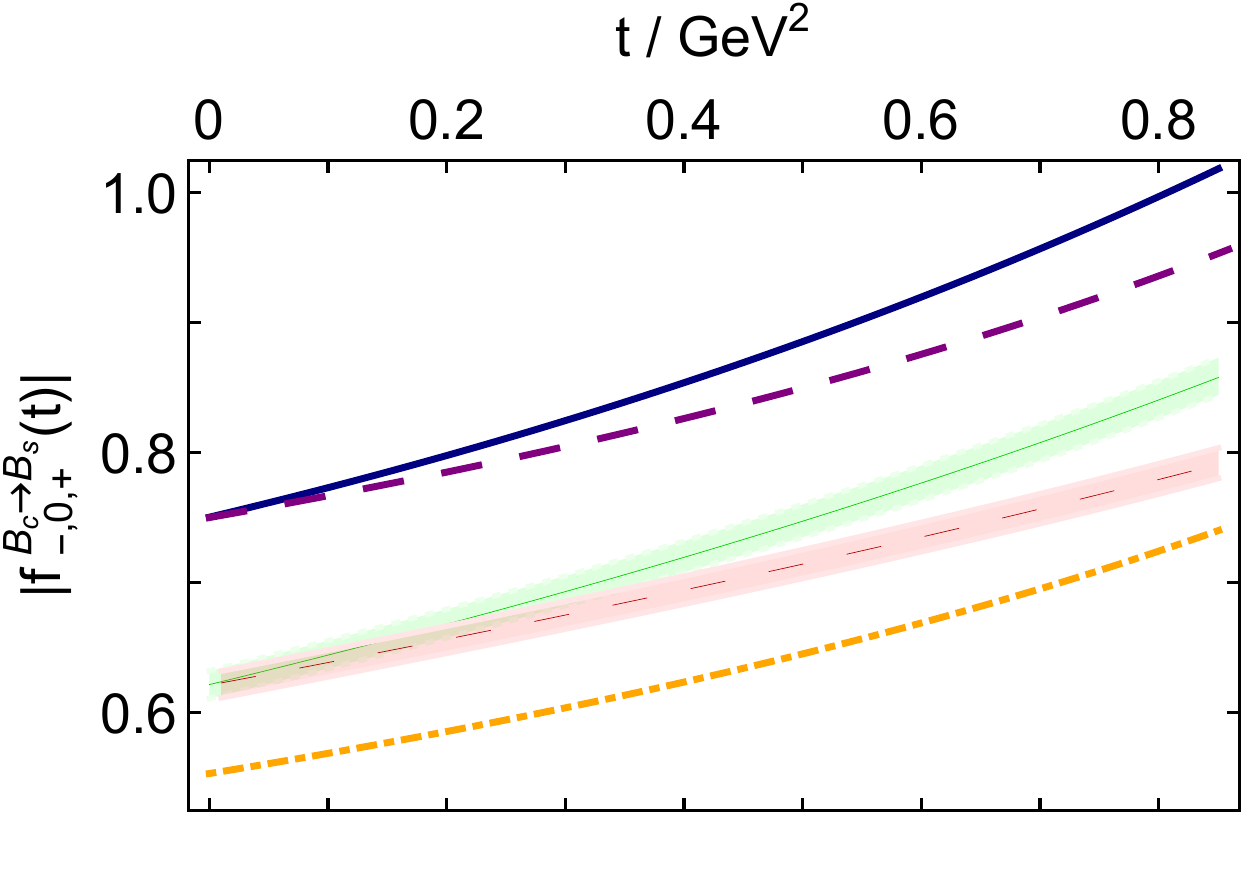}
\vspace*{-2ex}

\leftline{\hspace*{0.5em}{\large{\textsf{B}}}}
\vspace*{-4ex}
\includegraphics[width=0.45\textwidth]{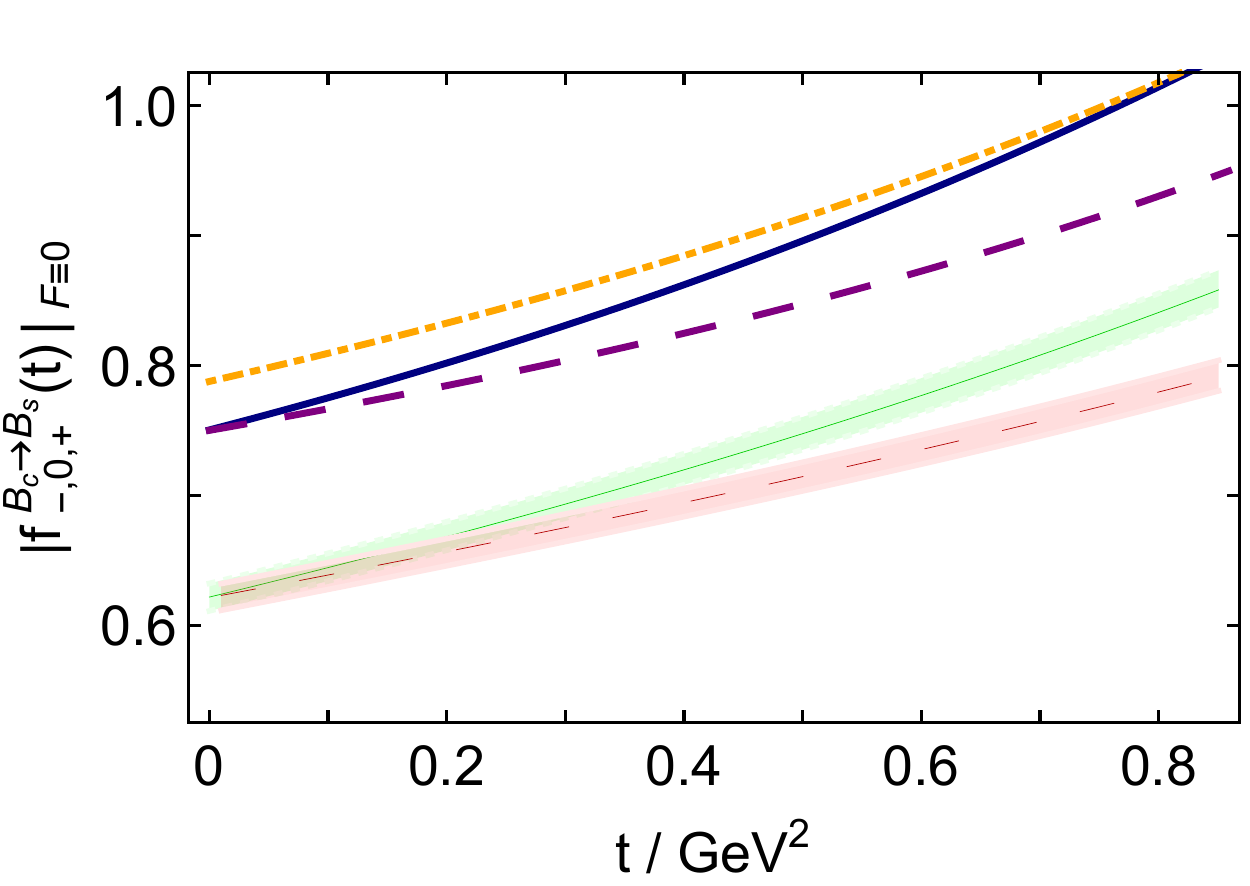}
\caption{\label{FigBcBs}
$B_c\to B_s $ transition form factors, defined by analogy with Eqs.\,\eqref{dMD}, \eqref{Projections}.
SCI results:
$f_+^{B_{b}^{s}}$, solid blue curve; $f_0^{B_{b}^{s}}$, long-dashed purple curve; and $-f_-^{B_{b}^{s}}$, dot-dashed orange curve.
\emph{Upper panel}\,--\,{\sf A} obtained with complete SCI Bethe-Salpeter amplitude in Eq.\,\eqref{PSBSA};
and \emph{lower panel}\,--\,{\sf B}, with $F_{0^-}\equiv 0$ and subsequent refitting.
In both panels, the comparison bands (thin dashed red, $f_0$; and thin green, $f_+$) are lQCD results from  Ref.\,\cite{Cooper:2020wnj}.
}
\end{figure}

Results for the maximum-recoil value of all seven of these transition form factors are listed in Table~\ref{fp0val}A.  The SCI and lQCD results are broadly in agreement, except for $B\to \pi$ and $B_s\to K$ (rows~4 and 5).  These last two transitions are special because of the enormous disparity in mass-scales between the initial and final states.  Surveying a raft of available model analyses \cite{Melikhov:1997wp, Melikhov:2001zv, Faessler:2002ut, Ebert:2003wc, Ball:2004ye, Wu:2006rd, Khodjamirian:2006st, Lu:2007sg, Ivanov:2007cw, Faustov:2014bxa}, one finds
\begin{equation}
f_+^{B\to \pi}(0) = 0.25(3), \; f_+^{B_s\to K}(0) = 0.32(4) ,
\end{equation}
values consistent with contemporary lQCD estimates.  We therefore judge that the SCI results in Table~\ref{fp0val}A are too large in these two cases.  Some failings had to be expected when attempting to develop an internally consistent SCI for use in treating systems with widely separated mass-scales.

Given these last observations, it is worth studying the $B\to \pi$ transition form factors in more detail.  Therefore, consider Fig.\,\ref{FigBpi}, which depicts SCI predictions for their $t$-dependence and comparisons with some lQCD results: despite the SCI's overestimate of the maximum recoil value of $f_+$, there is qualitative agreement over the entire physical $t$-domain.  Using the SCI result in Fig.\,\ref{FigBpi}A, one obtains the branching fraction listed in Table~\ref{fp0val}B, row~4: consistent with the overestimate of $f_+(0)$, the fraction is more than twice as large as the empirical value in Ref.\,\cite{Sibidanov:2013rkk}.  This indicates that the branching fraction integral is dominated by the transition form factor's behaviour on $0 < t \lesssim \tfrac{2}{3} t_m$.

$B_s\to K$ transition form factors are displayed in Fig.\,\ref{FigBsK}.  The comparison with quenched-lQCD results is qualitatively similar to that in Fig.\,\ref{FigBpi}.
Using the SCI result in Fig.\,\ref{FigBsK}A, one obtains the branching fraction listed in Table~\ref{fp0val}B, row~5.  Here, consistent with the overestimate of $f_+(0)$, the fraction is roughly twice as large as the model estimates in Refs.\,\cite{Wu:2006rd, Faustov:2014bxa}.

Regarding $B\to D$ semileptonic transitions, the mismatch between the masses of the initial and final states is much diminished as compared with the preceding two cases.  Consequently, the SCI results drawn in Fig.\,\ref{FigBD} match fairly well with existing data \cite{Aubert:2009ac} and recent parametrisations \cite{Yao:2019vty}.  (\emph{N.B}.\,Here and hereafter we draw $|f_-(t)|$ so as to optimise use of the plotting area.)  The SCI prediction for the branching fraction is listed in Table~\ref{fp0val}B, row~6, and aligns with experiment \cite[p.\,1531]{Zyla:2020zbs}.

We depict the SCI $B_s \to D_s$ transition form factors in Fig.\,\ref{FigBsDs}: they agree semiquantitatively with results from a recent lQCD study \cite{McLean:2019qcx}.  The SCI prediction for the branching fraction in Table~\ref{fp0val}B, row~7, is commensurate with estimates made using constituent quark models \cite{Zhao:2006at, Faustov:2014bxa}.

$B_c\to \eta_c$ transition form factors are plotted in Fig.\,\ref{FigBetac}.  In this instance, the only comparisons available are provided by exploratory lQCD calculations obtained with a highly improved staggered quark (HISQ) action \cite{Colquhoun:2016osw}: there are qualitative similarities between the results.  The SCI predictions for the branching fraction are listed in Table~\ref{fp0val}B, row~8.  The values agree with an estimate made using a hybrid scheme for combining perturbative- and lattice-QCD results \cite{Hu:2019qcn}.

\begin{figure}[t]
\includegraphics[width=0.45\textwidth]{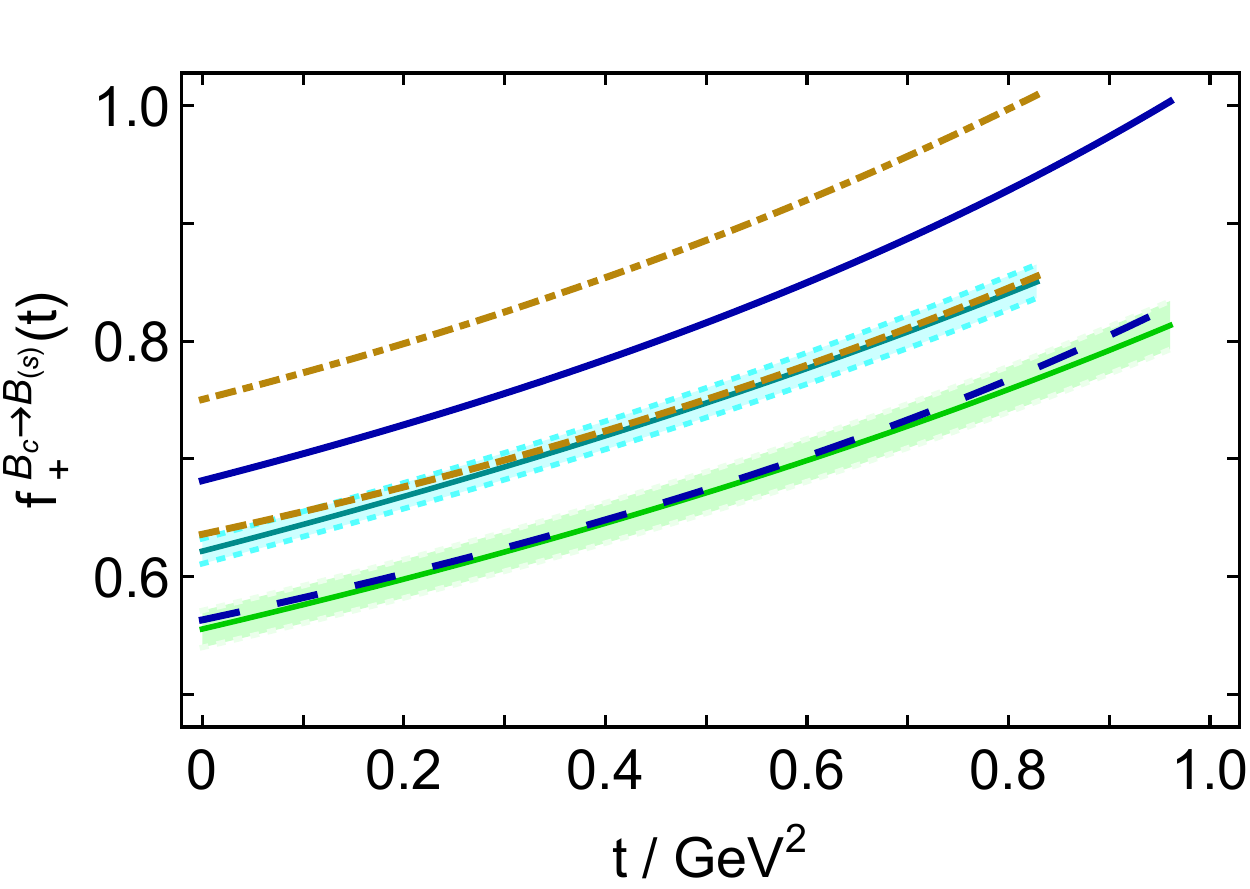}
\caption{\label{FigcfBcBBs}
SCI predictions for $f_+^{B_c\to B}$, solid blue curve, and $f_+^{B_c\to B_s}$, dot-dashed gold curve, compared with lQCD results for the same functions: $f_{+\,{\rm lQCD}}^{B_c\to B}$, thin green curve within like coloured band, and $f_{+\,{\rm lQCD}}^{B_c\to B_s}$, thin cyan curve within like coloured band.
In addition, calculated from the SCI results, $f_+^{B_c\to B}/1.21$, dashed blue curve, and $f_+^{B_c\to B_s}/1.18$, dot-dashed gold curve.
The comparison reveals that, in practical terms on the physical domains, the SCI and lQCD predictions differ only in their overall normalisations, \emph{i.e}.\ effectively, their respective values for $f_+(0)$.
}
\end{figure}

SCI predictions for the $B_c\to B$ form factors are drawn in Fig.\,\ref{FigBcB}, wherein they are compared with lQCD results obtained using $2+1+1$ flavours of dynamical sea quarks in the HISQ formalism \cite{Cooper:2020wnj}.  As highlighted by Fig.\,\ref{FigcfBcBBs}, the SCI and lQCD results for $f_{0,+}$ are qualitatively and semiquantitatively similar, with the SCI curves being approximately 21\% larger than those from the lQCD analysis, as measured using a ${\mathpzc L}_1$ measure.  The SCI prediction for the $B_c\to B$ branching fraction is given in Table~\ref{fp0val}B, row~9.  It fits comfortably within the wide range of model estimates listed in Ref.\,\cite{Ebert:2003wc}, \emph{viz}.\ $1.4(1.0)\times 10^{-3}$.
%
%The lQCD result in Ref.\,\cite{Cooper:2020wnj} is 32(10)\% larger than the SCI prediction.  This is puzzling, given that our results for $f_+^{B_c\to B}(0)$ are 21\% larger than the lQCD form factor, as highlighted by Fig.\ref{FigBcB}, and the branching fraction grows with $|f_+(t)|^2$ when all other quantities are kept fixed, Eq.\,\eqref{DecayFraction}.
%
Using the lQCD results drawn in Fig.\,\ref{FigBcB}, we obtain the $B_c\to B$ branching fraction listed in Table~\ref{fp0val}B, row~9, which is approximately 30\% smaller than the SCI prediction.

The final weak transition considered herein is $B_c \to B_s$, for which SCI predictions of the associated form factors are drawn in Fig.\,\ref{FigBcBs} and compared with lQCD results \cite{Cooper:2020wnj}.  As with $B_c\to B$, the SCI and lQCD results for $f_{0,+}$ are qualitatively and semiquantitatively similar: using a ${\mathpzc L}_1$ measure, the SCI curves are approximately 18\% larger than those from the lQCD analysis.  This is evident in Fig.\,\ref{FigcfBcBBs}.
The SCI prediction for the associated branching fraction is given in Table~\ref{fp0val}B, row~10.  Once again, it fits within the broad range of model estimates listed in Ref.\,\cite{Ebert:2003wc}, \emph{viz}.\, $16(11)\times 10^{-3}$.
%
%Similar to $B_c\to B$, the lQCD result in Ref.\,\cite{Cooper:2020wnj} is $49(7)$\% larger than the SCI prediction.  This is perplexing because our results for $f_+^{B_c\to B_s}(0)$ are 18\% larger than the lQCD form factor, as highlighted by Fig.\ref{FigBcBs}.
Using the lQCD results drawn in Fig.\,\ref{FigBcBs}, we obtain the $B_c\to B_s$ branching fraction listed in Table~\ref{fp0val}B, row~10, which is approximately 35\% smaller than the SCI prediction.

\section{Environment Sensitivity}
\label{ESensitivity}
%
% Vcd transitions ... D->pi  Ds-> K  Bc->B
% Vcs transitions ... D->K  Bc->Bs
% Vub  B->pi  Bs->K
% Vcb  B->D  Bs->Ds  Bc->etac
%
Regarding Nature's two mass generating mechanisms, it is interesting to consider the evolution of the form factors with increasing mass of the transition's inactive valence-quark, \emph{i.e}.\ as the current-quark mass produced by the Higgs boson becomes a more important part of the final-state meson's mass as compared with the EHM component.  Four classes can be formed from the semileptonic decays considered herein.
\begin{enumerate}[label=(\Roman*)]
\item $D\to\pi$, $D_s\to K$, $B_c\to B$: transition is $c\to d$; inactive valence-quarks are, respectively, $u$, $s$, $b$.
\item $D\to K$, $B_c\to B_s$: transition $c\to s$, inactive valence-quarks are $u$, $b$.
\item $B\to \pi$, $B_s \to K$: transition $b\to u$, inactive valence-quarks are $d$, $s$.
\item $B\to D$, $B_s\to D_s$, $B_c\to \eta_c$: transition $b\to c$, inactive valence-quarks are $u$, $s$, $c$.
\end{enumerate}

One may characterise this evolution by the values of $[1-f_+(0)]$ and $|f_-(0)|$ for the associated transitions: $[1-f_+(0)]=0=|f_-(0)|$ in systems constituted from mass-degenerate valence degrees-of-freedom.  The results can be read from Table~\ref{fp0val}A:
\begin{equation}
\begin{array}{lc|cccc}
     & & \rule{0.5em}{0ex} u/d & s & c & b \\ \hline
({\rm I})\ &1-f_+(0) \rule{0.5em}{0ex} & 0.24 & 0.25 & & 0.32 \\
& |f_-(0)| & 0.35 & 0.35 & & 0.49  \\\hline
({\rm II})\ &1-f_+(0) \rule{0.5em}{0ex} & 0.19 & & & 0.25 \\
& |f_-(0)| & 0.33 & & & 0.55\\\hline
({\rm III})\ &1-f_+(0) \rule{0.5em}{0ex} & 0.43 & 0.47 & & \\
& |f_-(0)| & 0.38 & 0.36 & & \\\hline
({\rm IV})\ &1-f_+(0)  \rule{0.5em}{0ex} & 0.25 & 0.26 & 0.27 & \\
& |f_-(0)| & 0.40 & 0.39 & 0.40 & \\\hline
\end{array}
\end{equation}
For comparison, analogous values for $\pi_{\ell 3}$ and $K_{\ell 3}$ transitions are, respectively, $(1.0,0.0)$ and $(0.98, 0.087)$ \cite{Chen:2012txa}.

The pattern of SCI results is clear: as the current-mass of the inactive valence degree-of-freedom increases, so do deviations from the flavour-symmetry limits.
The analysis in Ref.\,\cite{Yao:2020vef}, which used similar methods and a realistic quark+antiquark interaction, is too restricted in scope for a general comparison to be possible.  Nevertheless, the two Class (I) results are loosely compatible.
% f- increases D->pi ... Ds->K

Lattice-QCD results are available for all listed $f_+$ form factors, but they are typically from different collaborations.  (Class (IV) is something of an exception \cite{Cooper:2020wnj}.)  Reviewing those simulations, the pattern of lQCD results is nevertheless qualitatively similar to the SCI predictions.
However, this is not true for Class (III), \emph{viz}.\ the $B\to \pi$ and $B_s\to K$ transitions, which therefore demands further scrutiny.

\section{Isgur-Wise Function}
\label{Sec:IWF}
In considering semileptonic transitions connecting systems distinguished by a large disparity between the current-masses of their valence degrees-of-freedom, \emph{e.g}.\ heavy+light mesons, valuable simplifications can be exploited.  For instance, when $m_{Q_2} \gtrsim m_{Q_1} \gg m_q$, $H_{Q_2q} \to H^\prime_{Q_1 q}$ transitions between heavy+light pseudoscalar me\-sons are described by a single, universal function \cite{Isgur:1989ed}:
\begin{subequations}
\label{IWfn}
\begin{align}
f_\pm^{H\to H^\prime}(t) & =
\frac{ m_{H^\prime} \pm m_{H}}{2\sqrt{m_{H^\prime}m_{H}}}\, \xi(w(t))\,, \\
w(t) & = \frac{m_{H^\prime}^2+m_H^2 - t}{2  m_{H^\prime} m_{H}}\,.
\end{align}
\end{subequations}

\begin{figure}[t]
\includegraphics[width=0.45\textwidth]{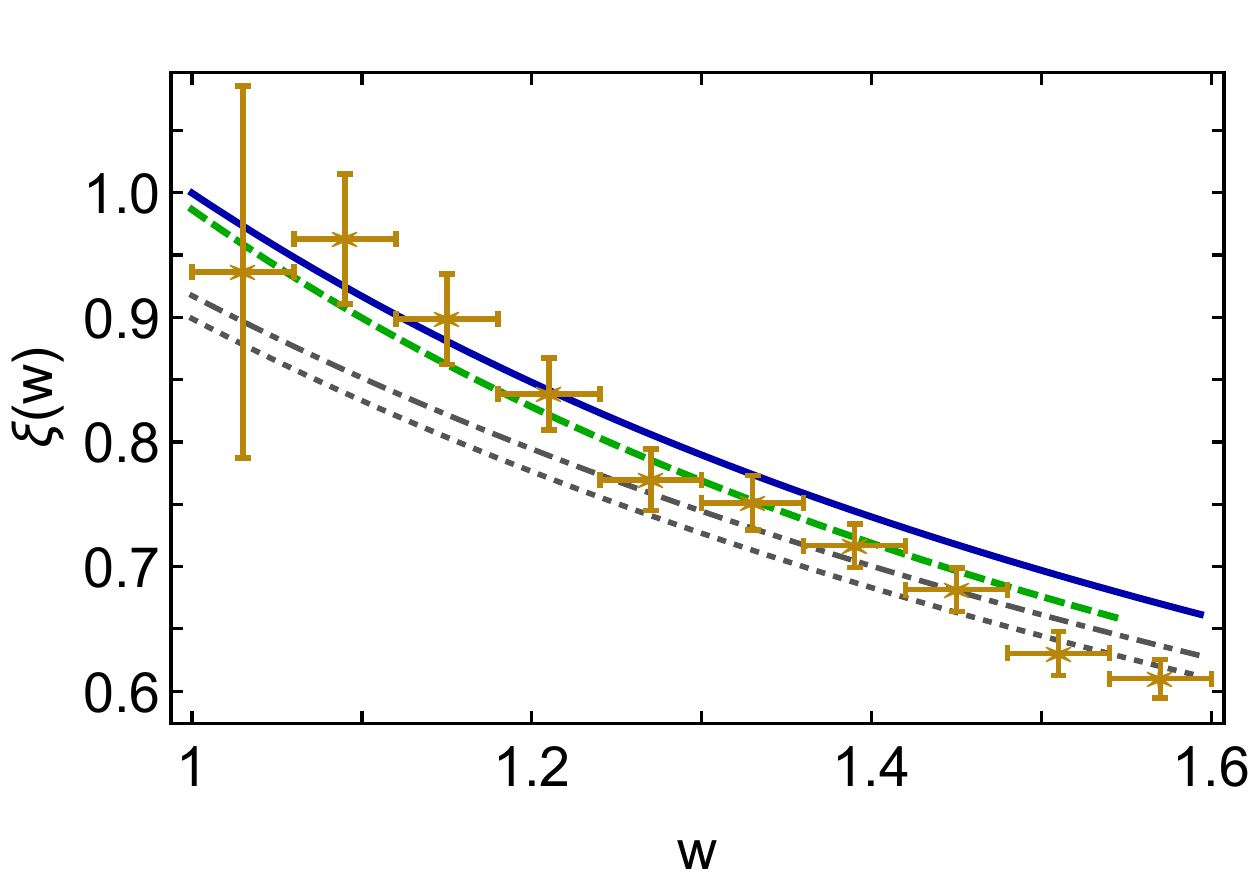}
\caption{\label{FigIW}
SCI predictions for the Isgur-Wise function, Eq.\,\eqref{IWfn}, as obtained from the $B\to D$ transition, solid blue curve, and the $B_s\to D_s$ transition, dashed green curve.
Data, gold stars, from Belle Collaboration measurement of $B\to D$ \cite{Glattauer:2015teq}.
Quark model results obtained from the $B\to D$ transition \cite{Choi:2021mni}: thin grey dot-dashed (linear potential) and dotted (harmonic oscillator) curves.
}
\end{figure}

Reviewing the transitions considered herein, two meet the requirements described above, viz.\ $B\to D$, $B_s\to D_s$; and in Fig.\,\ref{FigIW} we depict the function determined from those transition form factors using Eq.\,\eqref{IWfn}.  Evidently, the SCI produces results that conform with the notions of heavy-quark symmetry: the $B\to D$ and $B_s\to D_s$ results for $\xi(w)$ differ by $<1$\% according to a ${\mathpzc L}_1$ measure.  Moreover, the SCI prediction for the Isgur-Wise function, although somewhat stiff, is in fair agreement with empirical data \cite{Glattauer:2015teq}.

\section{Summary and Perspective}
\label{Epilogue}
We employed a symmetry-preserving regularisation of a vector$\times$vector contact interaction (SCI) to develop a unified treatment of the twelve independent semileptonic transitions involving $\pi$, $K$, $D_{(s)}$, $B_{(sc)}$ initial states and the masses and leptonic decays of the fifteen mesons that are either involved in the transitions or characterise the weak transition vertex.  The merits of this approach are its
algebraic simplicity;
paucity of parameters;
and simultaneous applicability to such a wide variety of systems and processes, sometimes involving large disparities in mass-scales between initial and final states.

Regarding meson masses, which are long-wavelength properties of the systems, the agreement between SCI predictions and experiment is good [Table~\ref{Tab:DressedQuarks}, Fig.\,\ref{FigMeson}], with comparison yielding a mean absolute relative difference $\overline{\rm ard} = 3$\%.  Concerning the leptonic decay constants, $\overline{\rm ard} = 13$\%: being dominated by ultraviolet momenta, the decay constants are a greater challenge for the SCI.

Turning to the $t$-dependence of transition form factors, the SCI predictions are often somewhat too stiff because the contact-interaction produces momentum-independent meson Bethe-Salpeter amplitudes.  Notwithstanding this, wherever experiment or sound theory results are available for comparison, the qualitative behaviour of the SCI results compares well, delivering semiquantitative agreement [Secs.\,\ref{SecSemiLepResults}, \ref{BWSLT}].  This is significant because all transitions are treated with the same three SCI-defining parameters.  The poorest comparison is found with $B\to\pi$ and $B_s\to K$.  These transitions, which have huge disparity between mass-scales of the initial and final states, present difficulties for all available methods.  Neglecting the $B\to \pi$ transition, comparisons between SCI predictions and experimentally measured branching fractions yield $\overline{\rm ard} = 7$\% [Table~\ref{fp0val}B].  Consequently, the SCI predictions for the five unmeasured branching fractions should be a sound guide.

An additional novel feature of our analysis is the information obtained about the effects that the Higgs mechanism of current-quark mass generation has on the form factors.  The suggestion is that as the current-mass of the non-transitioning valence degree-of-freedom increases, so do deviations from the flavour-symmetry limits [Sec.\,\ref{ESensitivity}].

Considering $B_{(s)}\to D_{(s)}$ transitions, we arrived at a prediction for the Isgur-Wise function [Sec.\,\ref{Sec:IWF}].  In comparison with recent $B\to D$ data from the Belle Collaboration, the SCI result produces $\chi^2/$datum$=1.9$.

Looking forward, it would be natural to use the SCI to unify the pseudoscalar-to-pseudoscalar transitions studied herein with pseudosca\-lar-to-vector semi\-leptonic transitions.  No parameters would need to be introduced or varied; and the results would provide additional benchmarks for analyses based on realistic momentum-dependent interactions.
Moreover, our analysis has highlighted a need for the use of symmetry-preserving formulations of realistic interactions in the study of all semileptonic transitions with $B_{(s,c)}$ initial states.  There is a dearth of such theory in this area; so with this in mind, the approach in Ref.\,\cite{Yao:2020vef} is currently being adapted to unify $(\pi,K,D,B)\to \pi$  and $(D_s,B_s)\to K$ transitions.

%Not a precision tool; but, can be used to identify consistency between results/predictions

\begin{acknowledgements}
We are grateful for constructive comments from S-.S.~Xu, Z.-Q.~Yao and P.-L.~Yin.
Work supported by:
National Natural Science Foundation of China (under Grant No.\,11805097);
Jiangsu Provincial Natural Science Foundation of China (under Grant No.\,BK20180323).
\end{acknowledgements}

%%\bibliographystyle{elsarticle-num-names}
%%\bibliography{../../../../CollectedBiB}

\end{document}